\documentclass[aps,prd,10pt,twocolumn,showpacs,amsmath,amssymb,nofootinbib,superscriptaddress]{revtex4-2}
\usepackage{graphicx}  
\usepackage{color}
\usepackage{times}
\usepackage{float}
\usepackage{afterpage}
\usepackage[utf8]{inputenc}
\usepackage{bm}
\usepackage{ulem}
\usepackage{multirow}
\usepackage{makecell}
\usepackage{url}
\usepackage{natbib}
\usepackage{mathrsfs}
\usepackage{physics}
\usepackage{comment}
\usepackage[table,xcdraw]{xcolor}
\usepackage{booktabs,makecell}
\usepackage{amsmath}
\usepackage{pifont}
\usepackage[colorlinks=true,citecolor=blue,urlcolor=blue,linkcolor=blue]{hyperref}
\usepackage{microtype}
\usepackage[none]{hyphenat}

\begin{document}

\title{Impact of dark matter distribution on neutron star properties}

\author{Ankit Kumar}
\email{ankitlatiyan25@gmail.com}
\affiliation{Department of Mathematics and Physics, Kochi University, Kochi, 780-8520, Japan}

\author{Hajime Sotani}
\affiliation{Department of Mathematics and Physics, Kochi University, Kochi, 780-8520, Japan}
\affiliation{Interdisciplinary Theoretical \& Mathematical Science Program (iTHEMS), RIKEN, Saitama 351-0198, Japan}
\affiliation{Theoretical Astrophysics, IAAT, University of T\"{u}bingen, 72076 T\"{u}bingen, Germany}

\date{\today}

\begin{abstract}
We investigate the structural and observable impacts of dark matter (DM) on neutron stars using a combined equation of state that integrates the relativistic mean field (RMF) model for baryonic matter with a variable density profile for DM, incorporating DM-baryon interactions mediated by the Higgs field. Employing three RMF parameter sets (NL3, BigApple, and IOPB-I) for baryonic matter, we analyze mass-radius relations, maximum mass, and tidal deformability, focusing on DM density scaling ($\alpha$) and steepness ($\beta$) parameters. Our findings reveal that increased DM concentration significantly enhances NS compactness, shifting mass-radius profiles and reducing tidal deformability. The DM influence strongly depends on the steepness of the DM density profile, where high $\beta$ values lead to strongly confined DM within the NS core, resulting in more compact and less deformable configurations. Observational constraints from PSR J0740+6620 and GW170817 impose consistent structural limits on DM fractions across different equations of state models, narrowing the allowable parameter space for DM and linking specific combinations of $\alpha M_{\chi}$ ($M_{\chi}$ being the mass of dark matter particle) and $\beta$ values to viable NS structures. This study highlights the interplay among DM concentration, nuclear stiffness, and observational data in shaping NS structure, offering insights into future constraints on DM in high-density astrophysical environments.
\end{abstract}


\maketitle

\section{Introduction}
\label{sec:1}
Neutron stars (NSs), the remnants of massive stellar explosions, are among the densest objects in the universe, providing a unique laboratory for studying matter at densities beyond nuclear saturation. Their interiors are thought to host a variety of exotic states of matter, including hyperons, di-baryons, nuclear pasta, and potentially free quarks~\cite{RevModPhys.80.1455, PhysRevC.95.055804, VIDANA2018112}. Understanding the equation of state (EoS) that describes the relationship between pressure and energy density in these environments is crucial for characterizing the mass-radius relationship of NSs, which can be mapped through the Tolman-Oppenheimer-Volkoff (TOV) equations~\cite{PhysRev.55.364, PhysRev.55.374}. Recent advances in observational techniques, such as the pulse profile modeling technique employed by NASA's Neutron Star Interior Composition Explorer (NICER), have improved our ability to infer the masses and radii of NSs, thus constraining the dense matter EoS \cite{Bogdanov_2019, Bogdanov_2021}. However, despite these advancements, significant uncertainties remain regarding the cold-dense matter EoS. 

A growing body of evidence suggests that dark matter (DM), which constitutes around 26.8\% of the universe's mass-energy content \cite{Jarosik_2011}, may play a significant role in the structure and evolution of NSs. The gravitational influence of DM, particularly in the form of weakly interacting massive particles (WIMPs), has been extensively studied as a potential contributor to the missing mass problem in galaxies and galaxy clusters~\cite{BERTONE2005279}. Although direct detection efforts have yet to identify a suitable DM candidate, NSs offer a promising avenue for indirect detection due to their extreme densities and strong gravitational fields. Their immense gravitational fields can capture DM particles, potentially leading to observable effects on the stars' properties. The capture and accumulation of DM within NSs can occur through elastic scattering with nucleons, with the dense stellar environment facilitating significant energy loss and efficient DM capture. Depending on the nature of the DM—whether it is self-annihilating or asymmetric—its presence can lead to a range of outcomes, including the formation of mini black holes or stable configurations within the star~\cite{PhysRevD.40.3221}. Studies have shown that the presence of DM can affect the mass-radius relationship of NSs, potentially leading to reductions in maximum attainable mass and shifts in the stellar radius. Such effects provide a unique opportunity to infer DM properties from astrophysical observations. As future missions such as STROBE-X and eXTP~\cite{ray2019strobexxraytimingspectroscopy, Watts2018} promise to deliver even more precise measurements of NS parameters, the integration of DM into NS models will become increasingly critical for advancing our understanding of both DM and the fundamental nature of matter at extreme densities. Moreover, the potential for DM to form compact objects, like boson stars or dark stars~\cite{PhysRev.172.1331, PhysRevLett.57.2485}, either independently or as part of a hybrid structure with NSs, presents a unique opportunity to explore DM properties through multi-messenger astronomy. This approach combines data from GW detections, x-ray observations, and radio surveys, enabling a comprehensive examination of DM's impact on astrophysical phenomena.

Traditionally, EoS models for NSs have focused exclusively on baryonic matter, considering the possible phase transitions and interactions of neutrons, protons, and other nuclear particles. However, the potential presence of DM necessitates re-evaluating these models to incorporate the effects of DM-baryon or DM self-interactions. Recent studies have explored various frameworks to model these interactions, including the two-fluid formalism, which treats DM and baryonic matter as distinct components that are coupled only gravitationally~\cite{PhysRevD.105.123034, particles7030040, PhysRevD.102.063028}. Alternatively, single-fluid models incorporate DM effects directly into the nuclear EoS, positing a particle interaction mechanism between standard model particles and DM particles, considering mediators to facilitate interactions between baryonic matter and DM~\cite{10.1093mnrasstad3658, PhysRevD.104.063028, CIARCELLUTI201119, PhysRevD.107.083522}. Theoretical models suggest that DM can exist in various forms, such as asymmetric or self-interacting bosonic and fermionic particles, which do not self-annihilate and may accumulate in significant quantities within NSs~\cite{PhysRevD.79.115016, PhysRevD.82.123512, 10.1093mnrasstad3658, Giangrandi_2023}. This accumulation can influence the star’s thermal and structural properties, potentially altering its gravitational wave signature during mergers~\cite{PhysRevD.103.123022, 10.1093/mnrasl/slv049, PhysRevD.99.043011}. DM can occupy two primary spatial configurations within and around NSs: 
a core integrated within its interior or 
a halo enveloping the star~\cite{ PhysRevD.105.023001}. Each configuration presents distinct implications for the star's gravitational mass, radius, and tidal deformability, which are critical parameters in the study of gravitational waves and NS mergers. The halo configuration suggests that DM forms an extended structure surrounding the NS, extending beyond the baryonic surface. This distribution can influence the star's external gravitational field, potentially altering its interaction with nearby matter and affecting its observable gravitational wave signals during mergers. The halo could also contribute to the overall mass, tidal deformability, and gravitational lensing effects attributed to the NS system~\cite{Nelson_2019, Giangrandi_2023}. Conversely, the core configuration involves a substantial concentration of DM within the NS itself. In this scenario, DM is integrated into the star's interior, contributing to the modification of internal pressure and energy density profiles. This integration can lead to changes in the star's EoS, which directly affect its mass-radius relationship, tidal deformability, and other properties such as thermal evolution and rotational dynamics~\cite{PhysRevD.97.123007, Konstantinou_2024}. It has also been hypothesized that DM cores could introduce distinct features in the gravitational wave spectrum post-merger, potentially creating additional peaks that differ from those generated by the baryonic components~\cite{ELLIS2018607}. These unique signatures could provide a new observational avenue for probing the presence and properties of DM within NSs.

Despite the extensive theoretical efforts to understand the role of DM in NSs, a key question that remains largely unresolved is the density distribution of DM particles within these dense stellar remnants. The impact of DM on NS observables, such as gravitational wave signatures, mass-radius profiles, and thermal evolution, is highly sensitive to its distribution within the star.  While various models have explored the potential effects of DM on NS properties, the actual distribution of DM within the star—whether it forms a uniform layer, is concentrated only in the core, or exhibits a gradient from core to crust—remains speculative. Studies addressing DM density distribution within NSs, often make simplifying assumptions (or typically involve constant DM density)~\cite{PhysRevD.84.107301, CIARCELLUTI201119, PhysRevC.89.025803}, which may not fully capture the complex interplay between DM and baryonic matter under existing extreme conditions. In the present work, we aim to address these gaps by investigating the density distribution of DM residing within the core of NSs, utilizing the DM admixed NS model introduced by Panotopoulos and Lopes~\cite{PhysRevD.96.083004}. While numerous investigations have utilized this model~\cite{PhysRevD.99.043016, PhysRevD.109.083021, PhysRevD.104.063028, Routaray_2023, Bhat2020, 10.1093/mnras/stac1013, ABAC2023101185, 10.1093/mnras/stab2387, PhysRevD.110.063001}, a critical area of study—the density distribution of dark matter within the core of NSs—remains unexplored. Previous studies have often assumed a constant DM Fermi momentum, leading to a uniform DM density throughout the star. However, this assumption appears physically implausible for several reasons. Firstly, the intense gravitational field of a NS suggests that if gravitationally trapped, DM should exhibit higher densities in the core compared to the crust or surface. Secondly, assuming a constant DM density can result in unphysical descriptions of the NS surface when integrating the TOV equations for a single-fluid model, particularly if phase transitions are not considered. Additionally, since in this model, we treat DM and baryonic matter as a single fluid without considering any phase transitions, observed deviations from universal relations in low-mass NSs with constant DM density are unexpected and suggest inconsistencies~\cite{PhysRevD.110.063001}. This work addresses these challenges by exploring a variable DM density profile within the NS core, factoring in gravitational influences, and ensuring alignment with the NS surface. This approach results in a more realistic depiction of DM distribution, providing novel insights into NS properties and establishing coherent constraints on the mass-density distribution of DM particles in light of the GW170817 data.

This paper is organized as follows: Section \ref{sec:2} presents the theoretical framework, including the relativistic mean-field (RMF) formalism for baryonic matter and the variable density profile for DM, alongside the coupled TOV equations used to calculate NS properties. In Sec. \ref{sec:3}, we analyze the results, beginning with the DM density distributions within NSs (Sec. \ref{sec:3a}) and their influence on mass-radius relations (Sec. \ref{sec:3b}). We then investigate the effects of DM on maximum mass, tidal deformability, and radius for 1.108 $M_{\odot}$ NS (Sec. \ref{sec:3c}), followed by constraints on DM parameter space derived from observational data (Sec. \ref{sec:3d}). In addition, Sec. \ref{sec:3e} examines the range of DM fractions across various NS models. Finally, we conclude in Sec. \ref{sec:4} with a summary of the main findings and implications for future research in DM admixed NSs.

\section{Formalism}
\label{sec:2}
Understanding the complex composition of DM admixed NSs relies on an EoS that integrates the contributions from both baryonic and DM components. In this work, we utilize the well-established Relativistic Mean Field (RMF) formalism to describe the baryonic sector, while the DM component is incorporated using a framework introduced by Panotopoulos and Lopes to characterize the DM contribution in NSs \cite{PhysRevD.96.083004}. These two components are combined into a single-fluid formalism, allowing us to examine the impact of DM on NS properties, such as mass-radius relations and tidal deformability, while adhering to constraints from recent observational data.

The baryonic EoS is derived from the RMF formalism, which effectively captures the balance between attractive and repulsive forces in nuclear matter, leading to a realistic description of EoS for baryonic matter \cite{MULLER1996508}. In this formalism, interactions between nucleons—primarily neutrons and protons— are mediated by meson exchange, specifically the scalar $\sigma$, vector $\omega$, and isovector $\rho$ mesons. The $\sigma$ meson facilitates the attractive interaction, while the $\omega$ meson provides the repulsive force necessary to achieve saturation in nuclear matter. The $\rho$ meson, on the other hand, contributes to the isospin asymmetry energy, which becomes significant at the high densities typical of NS interiors. The coupling constants between the mesons and baryons as well as the self and cross-coupling constants of mesons, are calibrated to match known nuclear properties at saturation density and measured properties of finite nuclei, ensuring consistency with the terrestrial nuclear experiments \cite{BOGUTA1977413, Gambhir1989, CHEN2015284, PhysRevC.90.044305}. 

In this work, we use three RMF parameter sets: NL3~\cite{PhysRevC.55.540}, BigApple~\cite{PhysRevC.102.065805}, and IOPB-I~\cite{PhysRevC.97.045806}. All three parameter sets reproduce the nuclear saturation properties, including saturation density, binding energy, and symmetry energy, in accordance with terrestrial nuclear experiments. However, while the IOPB-I and BigApple parameter sets also satisfy the nuclear incompressibility at saturation density, the NL3 parameter set does not fully meet this criterion \cite{WALECKA1974491, BOGUTA1977413}. Despite this limitation, NL3 is included in our analysis due to its ability to generate one of the stiffest EoS within the RMF framework, resulting in the highest predicted NS masses among the considered parameter sets. It is also regarded as a classical reference in RMF studies. For more in-depth information on the theoretical framework, mathematical expressions, and calibration of the RMF formalism, interested readers may review \cite{MULLER1996508, BOGUTA1977413, BOGUTA1977413, KUBIS1997191, PhysRevC.55.540, PhysRevC.62.015802, PhysRevC.90.044305, Kumar2020}.

Although separate models are employed for baryonic matter and DM, the Lagrangian of the adopted DM model includes an interaction term that indirectly influences the effective mass of nucleons within the RMF formalism. In this model, the Higgs field serves as a common mediator, facilitating separate interactions with both DM particles and nucleons. This indirect coupling alters the mesonic fields, leading to a redistribution of the scalar and vector fields within the RMF framework, thereby affecting the kinetics and mechanics of nucleons to some extent. The Lagrangian for the DM component which describes the dynamics of DM particles and their interaction with the Higgs field, is given by:
\begin{align}
    \mathcal{L}_{\rm DM} =& \,\bar\chi\left[i\gamma^\mu \partial_\mu - M_\chi + y h \right]\chi 
    \nonumber \\
    & + \frac{1}{2}\partial_\mu h \partial^\mu h - \frac{1}{2} M_h^2 h^2 + \frac{f\,M_{n}}{v} \bar \psi h \psi,   
    \label{eq:L_DM}
\end{align}
where $\chi$ represents the DM fermion field, $M_{\chi}$ is the DM particle mass, $h$ is the Higgs field, and $M_{h}$ denotes the Higgs mass. The coupling constant $y$ characterizes the interaction strength between the DM particles and the Higgs field, while $f M_{n}/v$ is the nucleon-Higgs coupling strength, with $M_{n}$ being the nucleon mass, and $v$ is the vacuum expectation value of the Higgs field.

The Lagrangian presented in Eq. \eqref{eq:L_DM} contains several parameters that need to be specified for modeling the DM-admixed NS system. For the coupling strength between DM and the Higgs field ($y$), we adopt a value of 0.06, which falls within the allowed range of 0.001-0.1 as per~\cite{PhysRevD.96.083004}. Notably, the system is not very sensitive to the exact value of $y$, primarily because the energy contribution of the term involving this coupling is typically minimal compared to the DM particle’s mass, which is in the GeV range \cite{PhysRevD.110.063001}. For the nucleon-Higgs coupling strength, $f M_{n}/v$, we use a value of $\approx$ $1.145 \times 10^{-3}$ \cite{PhysRevD.88.055025, PhysRevD.96.083004}, which reflects typical assumptions in the literature regarding nucleon-Higgs interactions.

A critical aspect of the DM component in NSs is the determination of its density distribution. In prior studies using this model, the DM density has generally been assumed to be constant throughout the star, corresponding to a constant Fermi momentum for DM particles. However, this assumption oversimplifies the complex dynamics within NSs and does not fully reflect the gravitational influence that should lead to a variable DM density profile. In this work, we adopt a simple but more realistic approach by defining the DM density as a function of baryon density, allowing for a more accurate depiction of DM distribution within the star's core. The DM density is defined as:
\begin{align}
   \frac{n_{\rm DM}}{n_{0}} = \alpha \Bigg(\frac{n_{\rm B}-n_t}{n_{0}}\Bigg)^{\beta},
   \label{eq:ndensity}
\end{align}
where $n_{\rm DM}$ is the DM number density, $n_{\rm B}$ is the baryon number density, $n_{0}$ is the nuclear saturation density without DM, $n_{\rm t}$ represents the core-crust transition density inside the NS, indicating the outer boundary of DM accumulation within the core, and $\alpha$ and $\beta$ are arbitrary dimensionless constants. The $\alpha$ and $\beta$ parameters in this equation are introduced to control the behavior and distribution of DM relative to the baryonic matter:
\begin{itemize}
    \item $\alpha$ represents the scaling factor for the DM density relative to nuclear saturation density. It effectively determines the relative abundance of DM at a given baryon density.
    \item $\beta$ governs the steepness of the DM density profile as a function of baryon density. A higher value of $\beta$ results in a sharper increase in DM density as the baryon density approaches the core region.
\end{itemize}
The choice of this variable DM density profile is motivated by the understanding that gravitational trapping of DM within NSs should lead to higher DM concentrations in the core compared to the outer layers. Since the DM is assumed to be confined within the core in this study, the DM density is set to zero in the region outside the core-crust transition density, i.e., $n_{\rm B} \le n_{\rm t}$, reflecting the assumption that DM particles are effectively trapped by the star’s gravitational field. In this work, we use the SLy4 EoS for the crust \cite{refId0}, and the values of the core-crust transition density, $n_{\rm t}$, for the three RMF parameter sets are as follows: $4.512 \times 10^{-2}$ $\rm fm^{-3}$ for NL3, $6.113 \times 10^{-2}$ $\rm fm^{-3}$ for BigApple, and $3.646 \times 10^{-2}$ $\rm fm^{-3}$ for IOPB-I. These transition densities define the maximum extent of DM confinement within the core region for each parameter set. This approach aligns with the notion that the DM component primarily resides within the denser central region of the NS, as suggested by prior theoretical and observational studies (as mentioned in section \ref{sec:1}). This variable DM density profile provides a more realistic model of DM distribution within NSs. Additionally, it facilitates a better examination of its effects on NS observables, such as mass-radius relations, tidal deformability, and gravitational wave signals. The impact of this distribution on the overall EoS of DM-admixed NSs will be analyzed in the subsequent section. 

\begin{figure*}[tbp]
    \centering
    \includegraphics[width=\textwidth]{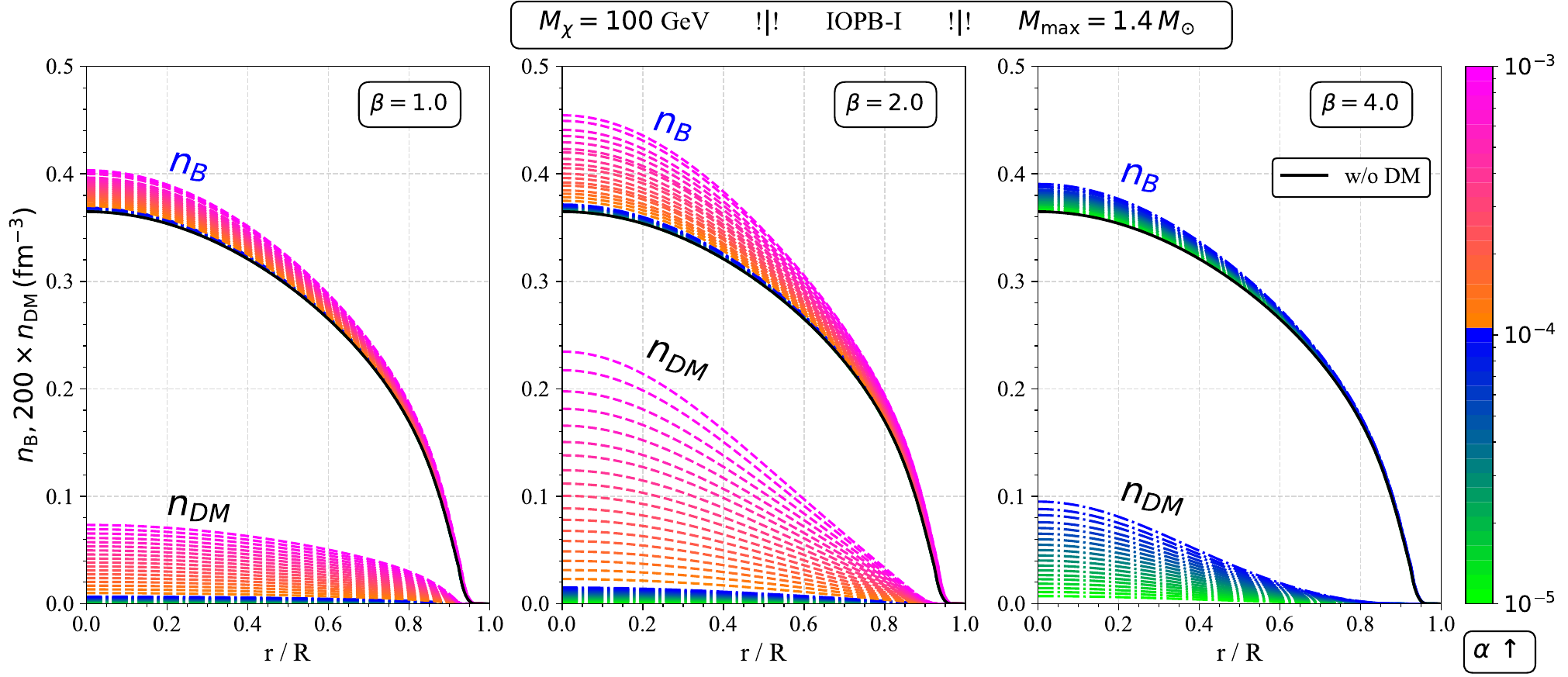}
    \caption{The radial profile of the DM number density, $n_{\rm DM}$, and number density of baryonic matter, $n_{\rm B}$, for $1.4M_\odot$ NS models constructed with the IOPB-I parameter set, varying the parameter $\alpha$ from $10^{-5}$ to $10^{-3}$ with $M_\chi=100$ GeV. The left, middle, and right panels correspond to the results with $\beta=1$, 2, and 4. In each panel, the solid line denotes the radial profile of $n_{\rm B}$ without DM. We note that since the maximum mass of NS for $\beta=4$ does not reach $1.4M_\odot$ with a larger value of $\alpha$ (see Fig.~\ref{fig:figure2}), we plot only with $\alpha=10^{-5}-10^{-4}$ for $\beta=4$.}
    \label{fig:figure1}
\end{figure*}

The energy density and pressure contributions for both baryonic matter ($\varepsilon_{\rm B}, P_{\rm B}$) and DM ($\varepsilon_{\rm DM}, P_{\rm DM}$) components are obtained by solving the energy-momentum tensor ($T^{\mu\nu}$), given as
\begin{eqnarray}
T^{\mu\nu} =  \sum_{j} \frac{\partial {\cal L}}{\partial(\partial_\mu\phi_j)}\partial^{\nu}\phi_j - \eta^{\mu\nu}{\cal L}, 
\end{eqnarray}
along with the self-consistent field equations for the mesons and the Higgs field within the RMF formalism, where $\phi_j$ includes all the fields present in the Lagrangian, ${\cal L}$. The EoS for the DM part is derived using the variable DM density profile introduced in Eq. \eqref{eq:ndensity}, incorporating the $\beta$ parameter ranging from 1 to 4, along with a specific range of $\alpha$ values. Additionally, while deriving the EoS, beta equilibrium and charge neutrality conditions are imposed to ensure that the system remains in chemical equilibrium, a necessary condition for the matter inside NSs. Once the EoS is established, it serves as the foundation for determining the mass-radius ($M$-$R$) profile of DM admixed NSs by solving the TOV equations \cite{PhysRev.55.364, PhysRev.55.374}, which describe the hydrostatic equilibrium of spherical, relativistic stars. The coupled TOV equations, which account for both baryonic and DM components, are given by:
\begin{eqnarray}
    \frac{dP(r)}{dr} &=& -\frac{ \left[ \varepsilon(r) + P(r) \right] \left[ m(r) + 4 \pi r^3 P(r) \right]}{r \left[ r - 2 m(r) \right]}, \\
    \frac{dm(r)}{dr} &=& 4 \pi r^2 \varepsilon(r),
\end{eqnarray}
where $P(r) = P_{\rm B} + P_{\rm DM}$ and $\varepsilon(r) = \varepsilon_{\rm B} + \varepsilon_{\rm DM}$ represent the total pressure and energy density at a radial distance $r$. The function $m(r)$ denotes the enclosed mass up to the radius $r$, which is calculated as the integration progresses outward from the center. The TOV equations are solved from the center to the surface of the neutron star, starting with initial conditions at the core: a central pressure $P_{c}$, determined by the EoS at the central energy density $\varepsilon_{c}$, and a mass function initialized as $m(0) = 0$. The integration continues until the pressure drops to zero, which signifies the star's surface. The total mass $M$ and radius $R$ are obtained at this boundary, providing the full mass-radius profile of the star. Solving the TOV equations allows for a detailed analysis of how the inclusion of DM affects the structural properties of NSs, including maximum mass, radius, and tidal deformability.

\begin{figure*}[tbp]
    \centering
    \includegraphics[width=\textwidth]{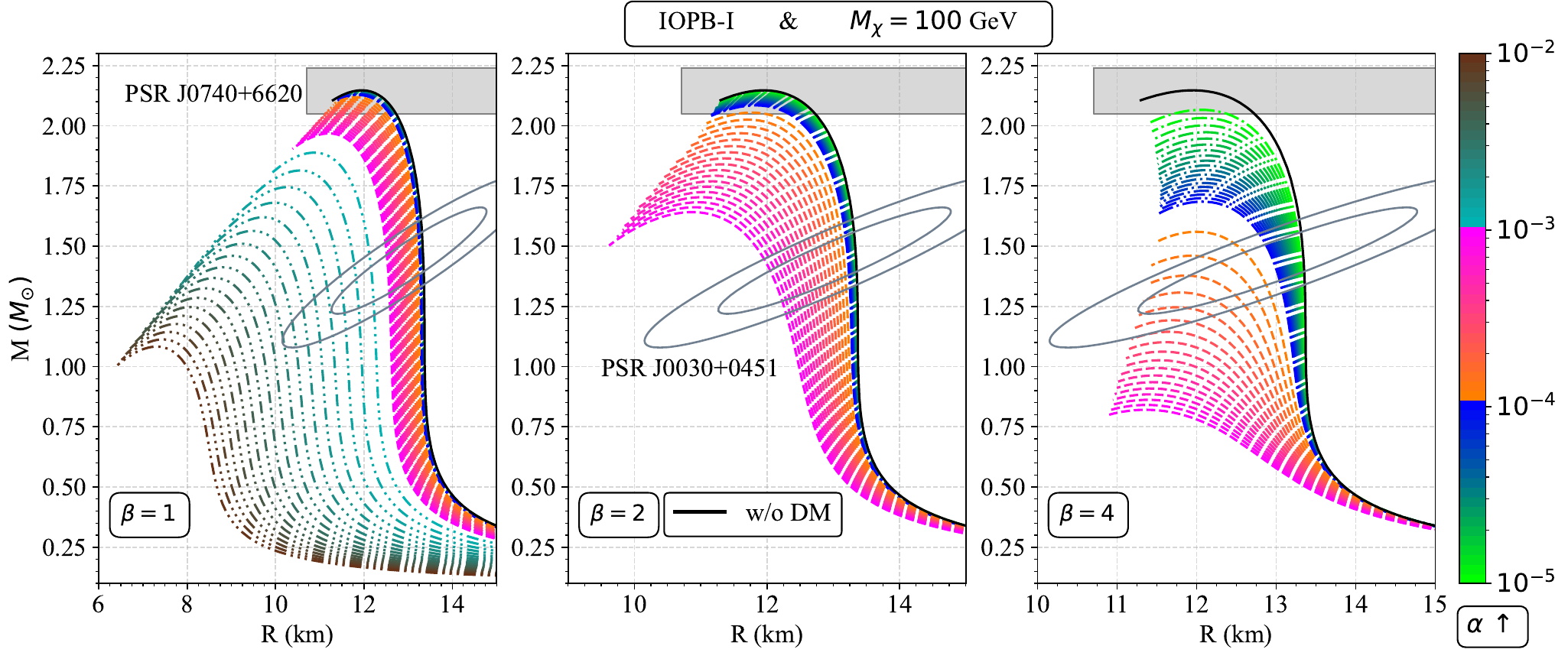}
    \caption{The mass and radius relations for the NS models constructed with the IOPB-I parameter set with $M_\chi=100$ GeV, adopting the radial profile of the DM number density given by Eq. (\ref{eq:ndensity}) with various values of $\alpha$. The left, middle, and right panels denote the results with $\beta=1,2$ and 4, respectively. In the figure, the value of $\alpha$ is varied from $10^{-5}$ to $10^{-2}$ for $\beta=1$ and from $10^{-5}$ to $10^{-3}$ for $\beta=2$ and 4. The NS mass and radius relation without DM are plotted with the solid black color line. For reference, the NS mass and radius constraints obtained by the NICER observation for PSR J0030+0451 and PSR J0740+6620 are shown with a double circle and with the shaded region (see text for details).}
    \label{fig:figure2}
\end{figure*}

\section{Results and Discussion}
\label{sec:3}
The results presented in this section investigate the structural properties of DM-admixed NSs, based on the formalism outlined earlier. The analysis primarily focuses on how varying DM density distributions influence the internal structure of NSs and extend to observable properties such as mass-radius relations and tidal deformability. As the discussion progresses, constraints on the DM parameter space are imposed using key observational data, including the gravitational wave event GW170817, the measured mass of PSR J0740+6620, and NICER observations of PSR J0030+0451. These constraints provide critical insights into how DM affects the dense matter EoS and its compatibility with current astrophysical observations, thereby refining our understanding of DM's role in NSs.

\subsection{Density distributions}
\label{sec:3a}

Figure~\ref{fig:figure1} presents the radial profiles of DM number density ($n_{\rm DM}$) and baryonic number density ($n_{\rm B}$) for $1.4 M_{\odot}$ NS models constructed using the IOPB-I parameter set. The DM profiles are shown for varying values of the scaling factor $\alpha$, ranging from $10^{-5}$ to $10^{-3}$, with a fixed DM particle mass $M_{\chi} = 100\, \rm GeV$. The left, middle, and right panels correspond to different values of the steepness parameter $\beta$, with $\beta = 1, 2, \mathrm{and}\, 4$, respectively. The solid line in each panel indicates the radial profile of $n_{\rm B}$ without DM, serving as a baseline for comparison. The DM density $n_{\rm DM}$ is multiplied by 200 to enhance its visibility, as indicated in the y-axis label. For $\beta = 4$, only lower values of $\alpha$ (up to $10^{-4}$) are shown, as higher values do not allow the model to achieve a 1.4 $M_{\odot}$ mass.

The radial profiles reveal that increasing $\alpha$ leads to a stronger concentration of DM within the NS while keeping the total NS mass constant at 1.4 $M_{\odot}$. This behavior aligns with the functional form defined in the formalism (Eq. \eqref{eq:ndensity}), where $\alpha$ governs the amplitude of DM density, while $\beta$ dictates the steepness. As $\alpha$ increases, $n_{\rm DM}$ becomes more prominent, peaking at approximately 0.33\% of $n_{\rm B}$ without DM for $\beta = 2$. This indicates that even a small DM fraction, relative to baryonic matter, can significantly alter the NS structure. For $\beta = 1$, the DM density spreads more uniformly across the core, while for $\beta = 4$, the distribution is sharply peaked at the center, reflecting strong central gravitational confinement.

The increased DM concentration enhances the gravitational potential within the NS core, necessitating a redistribution of baryonic matter to sustain hydrostatic equilibrium. As DM's gravitational attraction increases, baryons move inward to counter the enhanced core potential. This redistribution becomes evident at higher $\alpha$ values, with up to a 20\% increase in central $n_{\rm B}$ for $\alpha = 10^{-3}$ and $\beta = 2$, relative to the scenario without DM. This behavior demonstrates how baryonic matter adjusts to maintain stability under the stronger gravitational attraction of DM, while the total mass remains constant (see also Appendix \ref{sec:appendx1}).

It is important to emphasize that DM predominantly contributes to the energy density, rather than to the internal pressure of the NS. Consequently, gravitational confinement emerges as the primary mechanism by which DM modifies NS structure. Gravitational confinement refers to the phenomena where an increase in mass, in this case, due to added DM density, enhances the gravitational potential, attracting more matter, particularly baryons, toward the core. This compression of the NS core alters the baryonic distribution, setting the stage for the mass-radius ($M$-$R$) relations observed in Fig. \ref{fig:figure2}.

\subsection{Mass and radius relations}
\label{sec:3b}

Figure \ref{fig:figure2} illustrates the $M$-$R$ relations for NS models using the same IOPB-I parameter set and DM density profile as defined earlier, with $\alpha$ and $\beta$ varied within the specified ranges. The panels correspond to $\beta = 1, 2$, and 4 similar to Fig.~\ref{fig:figure1}, while the $\alpha$ values range from $10^{-5}$ to $10^{-2}$ for $\beta = 1$, and from $10^{-5}$ to $10^{-3}$ for $\beta = 2\, \mathrm{and}\, 4$. The solid line in each panel represents the $M$-$R$ relation without DM, serving as a reference. Observational constraints from NICER for PSR J0030+0451~\cite{Riley_2019, Miller_2019} and PSR J0740+6620~\cite{Riley_2021, Miller_2021} are depicted by a double circle and a shaded region, respectively. The tilted ellipses represent the $1\sigma$ (68\%) and $2\sigma$ (95\%) confidence regions for PSR J0030+0451 \cite{universe6060081}. For PSR J0030+0451, NICER primarily provides constraints on NS compactness, $M/R$, by carefully analyzing the pulsar’s light curve, which is highly sensitive to the gravitational field strength generated by the NS. In contrast, for PSR J0740+6620, additional mass measurements, specifically $M = 2.08 \pm 0.07 M_\odot$ \cite{Fonseca_2021}, enable NICER to provide radius constraints: $12.39^{+1.30}_{-0.98}$ km \cite{Riley_2021} and $13.7^{+2.6}_{-1.5}$ km \cite{Miller_2021}.

The $M$-$R$ relations reveal that increasing $\alpha$ consistently shifts NS models toward smaller radii for fixed masses, indicating increased compactness due to DM admixture. This effect becomes more pronounced at higher $\alpha$ values, consistent with the concentrated DM profiles observed in Fig.~\ref{fig:figure1}. As mentioned before, since DM does not contribute significantly to internal pressure, its presence primarily enhances gravitational confinement, compressing the NS further. Notably, steeper DM profiles, characterized by higher $\beta$ values (e.g., $\beta=2$ and 4), lead to even greater reductions in radii for corresponding masses (Fig. \ref{fig:figure2}), underscoring the central confinement induced by DM. The role of $\beta$ is not limited to controlling the DM distribution's steepness; it also modulates the degree of core compression. Higher $\beta$ values intensify central confinement, reflecting a more concentrated DM mass within the core, which directly correlates with sharper reductions in radii for a given NS mass.

\begin{figure*}[tbp]
    \centering
    \includegraphics[width=\textwidth]{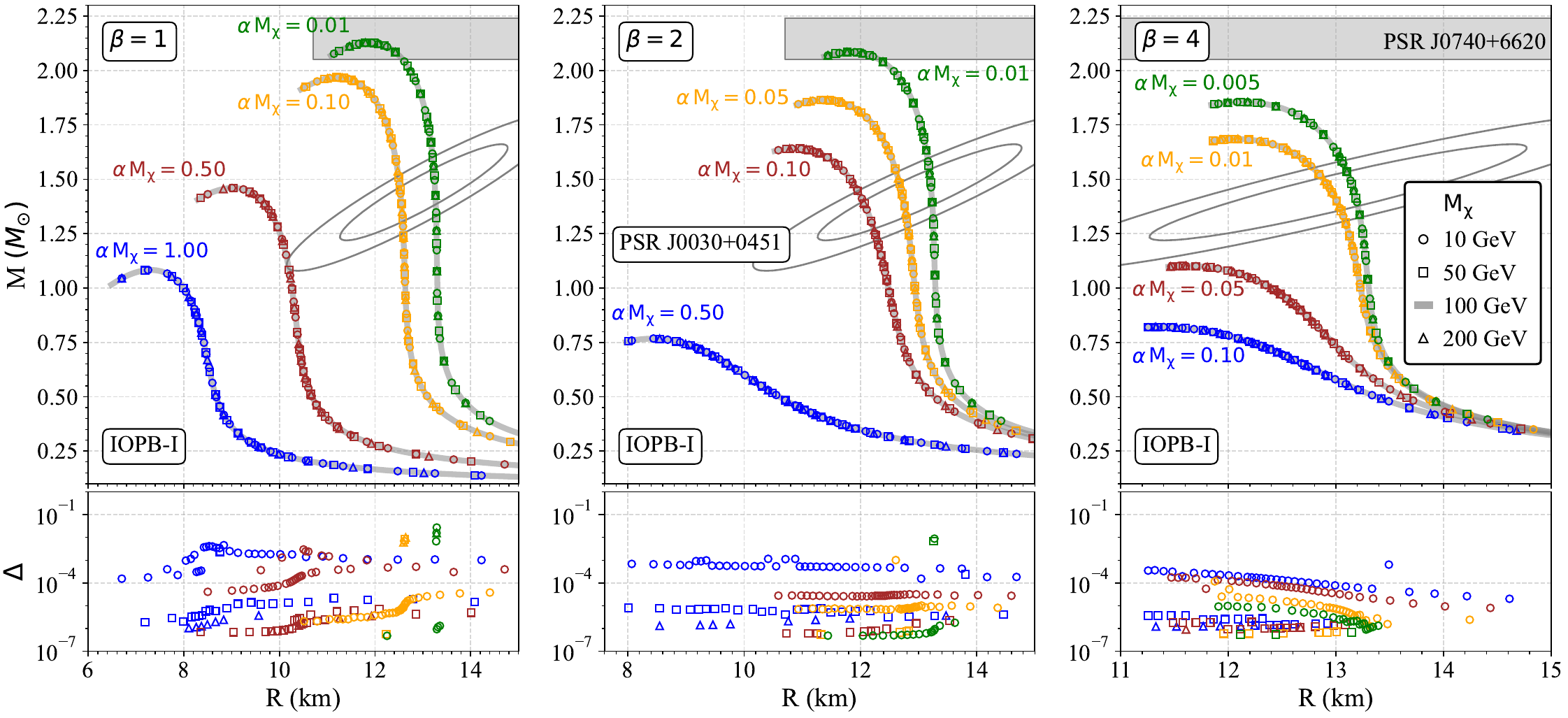}
    \caption{The NS mass and radius relation for fixed values of $\alpha M_\chi$ in the unit of GeV. The left, middle, and right panels correspond to the results for $\beta=1$, 2, and 4, adopting the IOPB-I parameter set. In each panel, the circles, squares, and triangles denote the NS models with $M_\chi=10$, 50, and 200 GeV, respectively, while the solid line denotes the NS models with $M_\chi=100$ GeV. The relative deviation $\Delta$ calculated with Eq.~(\ref{eq:Delta}) is shown in the bottom panels. For reference, the NS mass and radius constraints obtained by the NICER observation for PSR J0030+0451 and PSR J0740+6620 are shown with a double circle and shaded region respectively.}
    \label{fig:figure3}
\end{figure*}

It is also notable that while increased DM concentration leads to more compact NS configurations, some $M$-$R$ relations still overlap with NICER’s observational constraints. Lower $\alpha$ values, particularly for $\beta = 1$, produce $M$-$R$ curves that lie within NICER’s constraints for PSR J0030+0451 and PSR J0740+6620, indicating that modest DM admixture can yield physically consistent NS models. However, for higher $\beta$ and $\alpha$ values, the $M$-$R$ relations significantly deviate from observed limits, suggesting that highly concentrated DM distributions may result in overly compact configurations that are inconsistent with observed NS masses and radii.

An intriguing observation emerges from Fig.~\ref{fig:figure3}: we find that the $M$-$R$ profiles for DM admixed NSs remain almost identical when the product $\alpha M_{\chi}$ is kept constant, regardless of individual $\alpha$ or $M_{\chi}$ values, provided that $\beta$ is fixed. For instance, the $M$-$R$ curve for $M_{\chi} = 100$ GeV and $\alpha = 10^{-4}$ is almost the same as that for $M_{\chi} = 10$ GeV and $\alpha = 10^{-5}$ across all values of $\beta$. This behavior suggests that the impact of DM on NS compactness is governed primarily by its total energy density ($\varepsilon_{\rm DM} \propto \alpha M_\chi$), rather than by $\alpha$ or $M_{\chi}$ individually. The combined influence of $\alpha$ and $M_{\chi}$, through their product $\alpha M_{\chi}$, plays a direct role in determining the total energy density within the core. 

This observation is further supported by the lower panels of Fig. \ref{fig:figure3}, which depict the relative deviation $\Delta$ of the NS mass at a given radius $R$, denoted as $M(R)$, from the NS mass for the same radius with $M_{\chi} = 100$ GeV, denoted as $M_{100}(R)$, while keeping $\alpha M_{\chi}$ constant. This deviation is mathematically defined as
\begin{equation}
  \Delta = \frac{|M(R) - M_{100}(R)|}{M_{100}(R)}.
  \label{eq:Delta}
\end{equation}
The deviation $\Delta$ remains negligible across different $M_{\chi}$ values, indicating that the $M$-$R$ profiles are virtually unchanged for a constant $\alpha M_{\chi}$. This consistency underscores the gravitational effects discussed earlier, where the DM’s contribution to the gravitational field induces the redistribution of baryons to maintain hydrostatic equilibrium. Physically, this emphasizes that gravitational confinement in NSs is primarily determined by the combined effect of DM mass and concentration, as captured by the product $\alpha M_{\chi}$.

These findings have significant implications for gravitational wave observations, as the compactness of NSs affects their tidal deformability. The tidal deformability parameter, which influences the gravitational wave signal during NS mergers, may also be governed by the total gravitational energy represented by $\alpha M_{\chi}$. Observations of NS mergers could therefore provide important insights into the presence of DM, especially by focusing on deviations in tidal deformability linked to the $\alpha M_{\chi}$ contribution.

For future theoretical models, this suggests that parameter spaces exploring DM influences should prioritize combinations of $\alpha$ and $M_{\chi}$ that reflect the total energy contribution, rather than treating them independently. This approach will provide a more realistic representation of DM effects on NS structure and observable properties. In particular, models aiming to incorporate DM's influence on NSs should focus on the combined effect of DM concentration and particle mass, as encapsulated by $\alpha M_{\chi}$, to accurately evaluate its impact on NS observables and behavior.

\subsection{Maximum mass and tidal deformability}
\label{sec:3c}

\begin{figure}[tbp]
    \centering
    \includegraphics[width=\columnwidth]{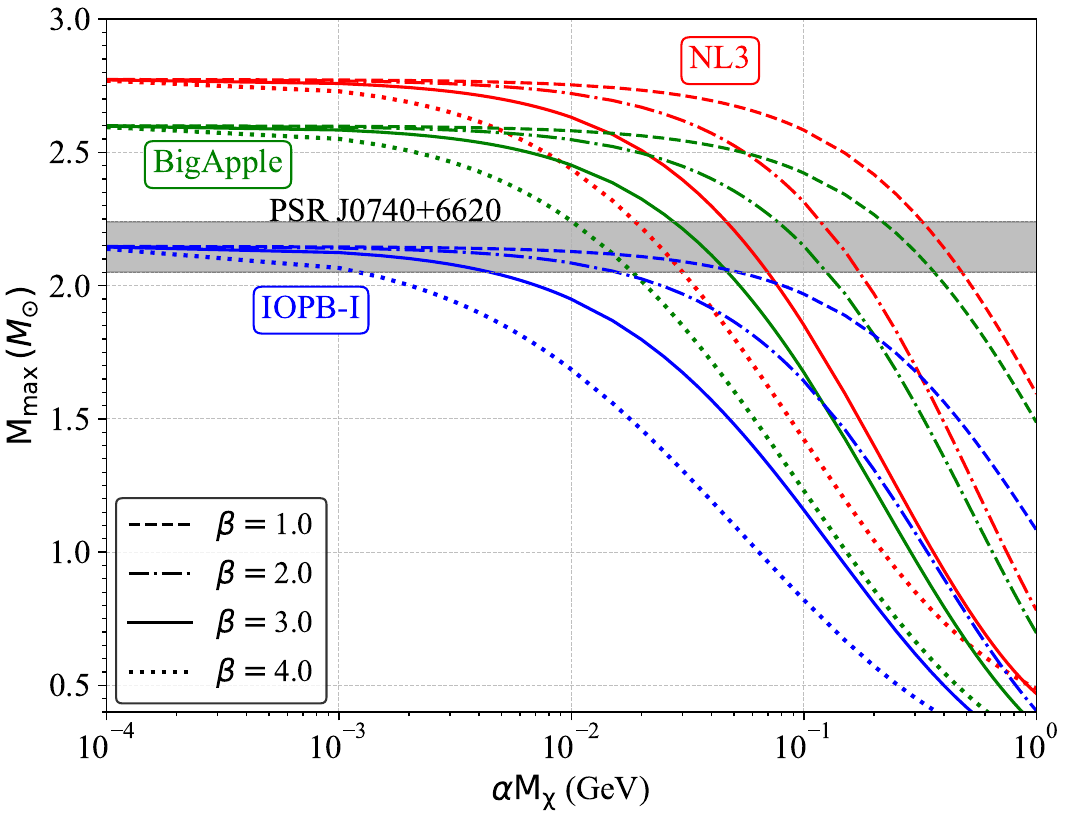}
    \caption{The NS maximum mass dependence on $\alpha M_\chi$, adopting the NL3, BigApple, and IOPB-I parameter sets. The dashed, dot-dashed, solid, and dotted lines correspond to the results for $\beta=1$, 2, 3, and 4, respectively. For reference, we also show the mass of PSR J0740+6620, which is $M=2.08^{+0.07}_{-0.07}M_\odot$ \cite{Fonseca_2021}. 
    }
    \label{fig:figure4}
\end{figure}

The identical $M$-$R$ profiles observed for a constant product $\alpha M_{\chi}$ provide a crucial insight into the dominant influence of DM energy density on NS compactness. We now extend our analysis to explore how this product affects other NS properties, such as the maximum mass and tidal deformability. Figure \ref{fig:figure4} illustrates the variation of the NS maximum mass as a function of the product $\alpha M_{\chi}$, employing three different nuclear EoS parameter sets: NL3, BigApple, and IOPB-I. The lines in the plot correspond to various values of the steepness parameter, $\beta$, with $\beta = 1, 2, 3, 4$ represented by dashed, dot-dashed, solid, and dotted lines, respectively. We also indicate the observed mass of PSR J0740+6620, which provides an upper and lower limit of $M=2.08^{+0.07}_{-0.07} M_\odot$ as reported in \cite{Fonseca_2021}. The models shown in this figure are in practice constructed by varying the value of $\alpha$ from $10^{-6}$ to $10^{-2}$ with a fixed DM particle mass $M_\chi = 100$ GeV, which allows us to investigate a range of $\alpha M_\chi$ values spanning from $10^{-4}$ to $10^{0}$ GeV.

The trends in Fig. \ref{fig:figure4} reveal that the maximum mass of NS decreases consistently as the product $\alpha M_\chi$ increases across all EoS models and $\beta$ values. This reduction in maximum mass is consistent with the earlier discussions of gravitational confinement - adding more DM energy density increases the gravitational potential, resulting in more compact and, consequently, less massive NS configurations. This trend is observed irrespective of the specific EoS, highlighting the broad applicability of the conclusions across different nuclear matter models. Interestingly, while all models show a similar qualitative trend, the maximum mass reaches different quantitative values, depending on the EoS used. This indicates that while DM energy density is a critical factor, the underlying nuclear interactions described by the EoS also play a significant role in determining the ultimate stability of NS configurations.

\begin{figure}[tbp]
    \centering
    \includegraphics[width=\columnwidth]{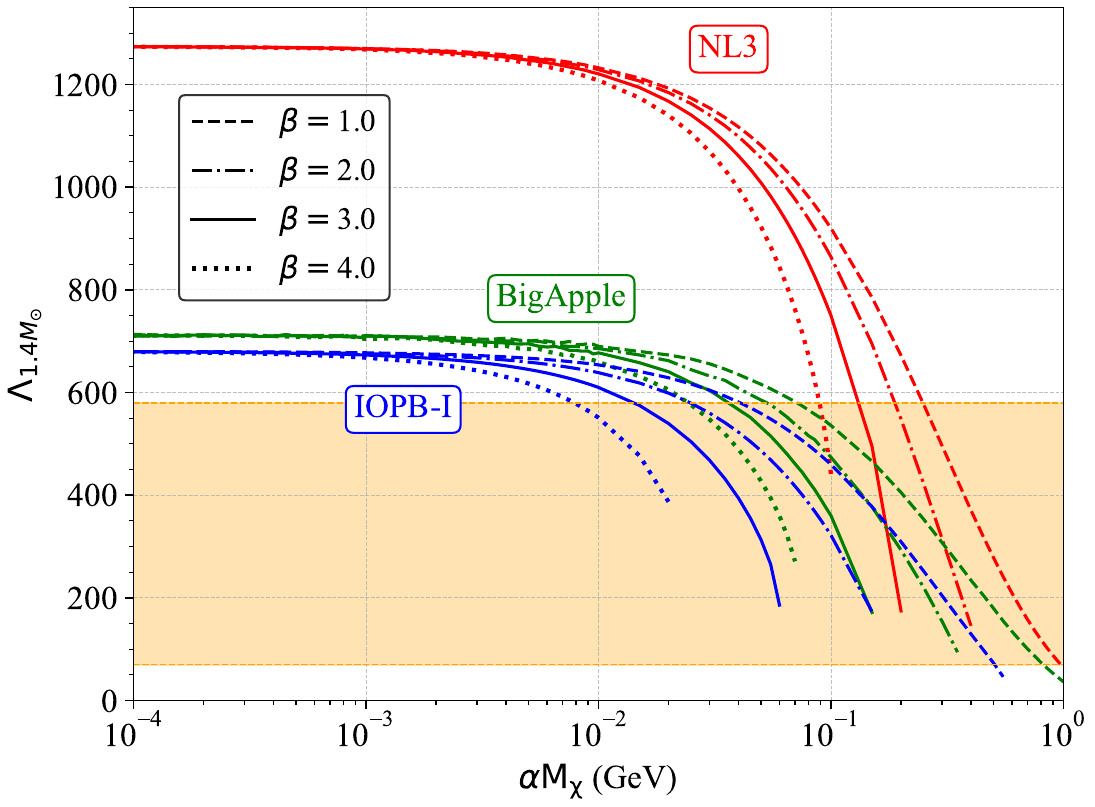}
    \caption{Same as Fig.~\ref{fig:figure4}, but for the dimensionless tidal deformability for the $1.4M_\odot$ NS model, $\Lambda_{1.4M_\odot}$. For reference, we show the constraints on $\Lambda_{1.4M_\odot}$ obtained from the GW170817, i.e., $\Lambda_{1.4M_\odot}=190^{+390}_{-120}$ \cite{PhysRevLett.121.161101}, with the shaded region.}
    \label{fig:figure5}
\end{figure}

The dependence of the maximum mass on $\alpha M_\chi$ also highlights the role of the steepness parameter $\beta$. For $\beta = 1$ (dashed lines), the reduction in maximum mass is more gradual compared to steeper profiles ($\beta = 3, 4$). This behavior can be attributed to the degree of central concentration induced by higher $\beta$ values. Specifically, higher $\beta$ values imply a more concentrated DM profile, which further enhances gravitational confinement, leading to a steeper decline in maximum mass (see also Appendix \ref{sec:appendx1}). For $\beta = 4$, the maximum mass drops sharply, demonstrating that increased central confinement due to DM leads to stronger compression and ultimately a lower mass limit.

This trend emphasizes the importance of balancing DM mass and concentration in models that aim to satisfy observed NS properties. While an increase in $\alpha M_\chi$ leads to enhanced gravitational potential and more compact configurations, it comes at the cost of reducing the overall stability and mass-supporting capability of the NS. The observational constraint from PSR J0740+6620 serves as a useful benchmark in this analysis. For relatively low $\alpha M_\chi$ values, the predicted maximum masses for all parameter sets remain above the observed mass, suggesting that modest DM admixtures are compatible with known astrophysical observations. However, as $\alpha M_\chi$ increases, the models with higher $\beta$ increasingly deviate, dropping below the observational limit. This indicates that highly concentrated DM cores, particularly those characterized by high $\beta$ and large $\alpha M_\chi$, are less likely to correspond to physical NS configurations consistent with observed masses.

Continuing with this line of analysis, Fig. \ref{fig:figure5} examines the effect of DM on the tidal deformability ($\Lambda_{1.4}$) for $1.4 M_{\odot}$ NSs, using the same parameter sets as those in Fig. \ref{fig:figure4}. The tidal deformability parameter is crucial as it characterizes how easily a NS deforms under the influence of tidal forces, such as those experienced in a binary merger scenario. Observational constraints from the gravitational wave event GW170817 are also shown, providing an important benchmark for comparing our model results. The trends in tidal deformability share similarities with the maximum mass trends—an increasing $\alpha M_\chi$ leads to a significant reduction in $\Lambda_{1.4}$ across all EoS models. This reduction reflects the increased compactness of NSs with higher DM admixture, leading to less deformable structures under tidal forces.

\begin{figure}[tbp]
    \centering
    \includegraphics[width=\columnwidth]{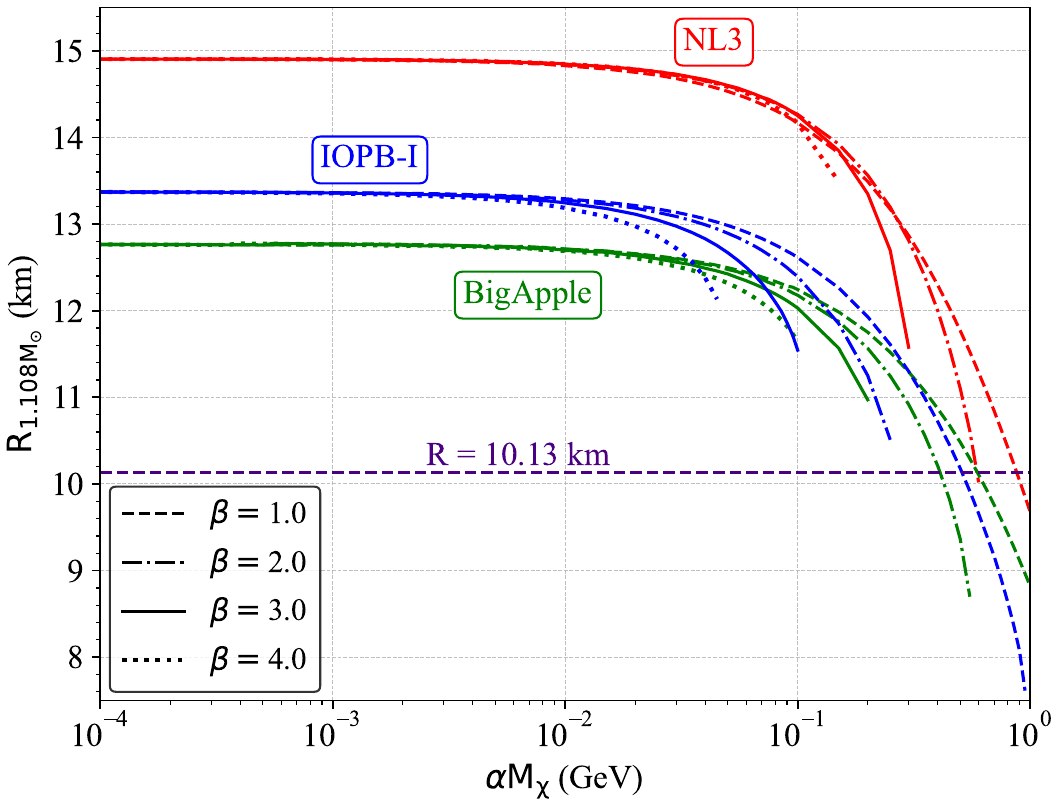}
    \caption{Same as Fig.~\ref{fig:figure4}, but for the radius with $M=1.108M_\odot$ NS models, $R_{1.108M_\odot}$. From the NICER observation for PSR J0030+0451, $R_{1.108M_\odot}$ should be larger than 10.13 km}
    \label{fig:figure6}
\end{figure}

A closer comparison of Figs. \ref{fig:figure4} and \ref{fig:figure5} reveal that the impact of $\alpha M_\chi$ is to consistently push both the maximum mass and tidal deformability towards values that are more compact and less deformable. However, a crucial observation is that while the maximum mass sets an upper limit based on stability, the tidal deformability results provide complementary constraints by linking the NS's internal compactness to observable gravitational wave signatures. For $\beta = 1$ and lower $\alpha M_\chi$ values, the predicted tidal deformability lies well within the range observed by GW170817, suggesting that moderate DM admixture is compatible with current observational limits. For higher values of $\beta$ and $\alpha M_\chi$, the tidal deformability falls below the lower observational limit, indicating configurations that are overly compact to be consistent with observed merger events. This dual analysis of maximum mass and tidal deformability, both influenced by $\alpha M_\chi$, provides a more comprehensive understanding of how DM affects NS properties and places stringent constraints on the parameter space of DM models.

Building upon the insights gained from the maximum mass and tidal deformability trends, we turn our attention to another key observable: the radius of the NS with $M = 1.108 M_\odot$, denoted as $R_{1.108M_\odot}$, shown in Fig. \ref{fig:figure6}. The plot illustrates the dependence of $R_{1.108M_\odot}$ on the product $\alpha M_\chi$, using the NL3, BigApple, and IOPB-I parameter sets. Similar to previous figures, the dashed, dot-dashed, solid, and dotted lines correspond to $\beta = 1, 2, 3$, and $4$, respectively, and the NICER observation for PSR J0030+0451 provides a constraint on $R_{1.108M_\odot}$, i.e., $R_{1.108M_\odot} > 10.13$ km.

Notably, Fig. \ref{fig:figure6} indicates that the radius for the $1.108 M_\odot$ NS models remains consistently above this lower observational limit for nearly all parameter combinations of $\alpha M_\chi$ explored, regardless of the RMF parameter set or steepness parameter $\beta$. The influence of increasing $\alpha M_\chi$ on $R_{1.108M_\odot}$ remains modest, with variations that do not strongly affect the agreement with observational constraints. This suggests that, unlike the more stringent limitations imposed by maximum mass and tidal deformability, the radius constraint for PSR J0030+0451 is relatively easily satisfied across a broad range of DM properties. The relative insensitivity of $R_{1.108M_\odot}$ to changes in $\alpha M_\chi$ indicates that, while DM admixture certainly modifies the overall NS structure, its influence on the radius for the $1.108 M_\odot$ NS models is less dramatic compared to the effects on maximum mass or tidal deformability. The NICER constraint on $R_{1.108M_\odot}$ thus acts as a relatively loose benchmark, highlighting that while DM admixture affects the NS structure, it does not strongly constrain the parameter space through this observable alone, unlike the more definitive constraints obtained from other observables like tidal deformability and maximum mass.

\subsection{Observational constraints on the DM parameter space}
\label{sec:3d}

\begin{figure}[tbp]
    \centering
    \includegraphics[width=\columnwidth]{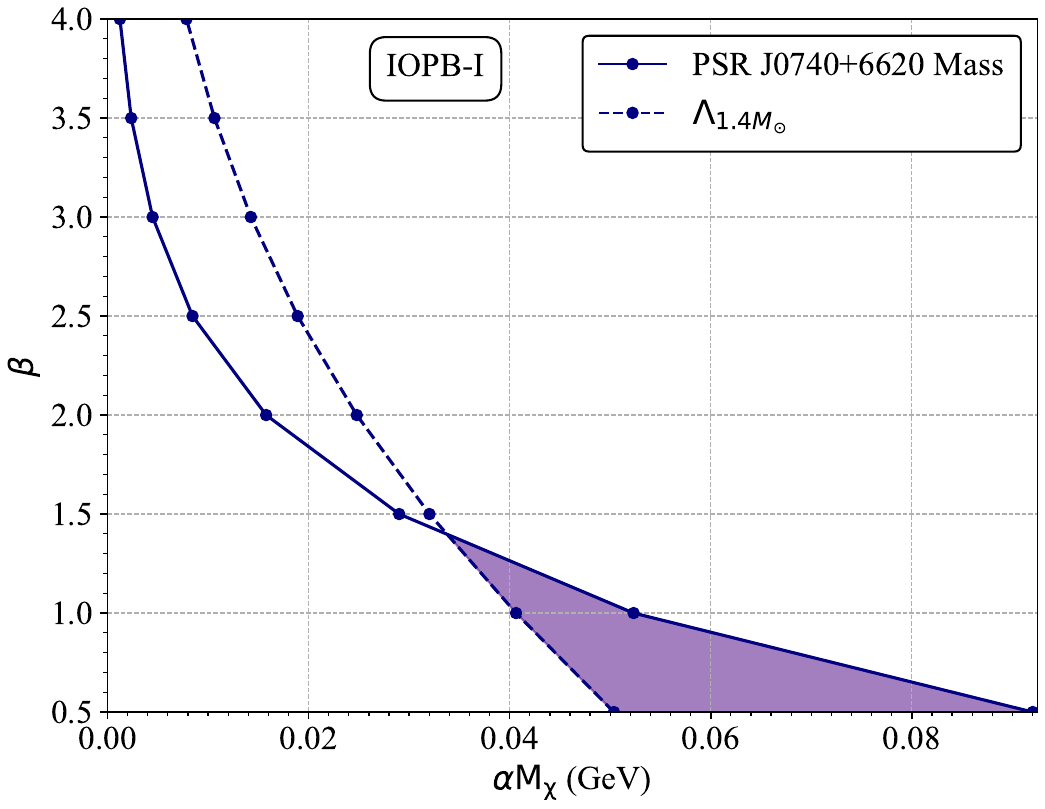}
    \caption{Allowed region in the parameter space of $\alpha M_\chi$ and $\beta$ obtained from the astronomical observations for PSR J0740+6620 and the GW170817 event, assuming the IOPB-I parameter set. The mass of PSR J0740+6620 gives us the upper limit (solid line), while the GW170817 event gives us the lower limit (dashed line).}
    \label{fig:figure7}
\end{figure}

Building on the preceding analyses of maximum mass and tidal deformability, Fig. \ref{fig:figure7} illustrates the allowed parameter space for $\beta$ versus $\alpha M_\chi$, focusing specifically on the IOPB-I parameter set. This plot consolidates the constraints derived from both the PSR J0740+6620 mass limit and the tidal deformability obtained from GW170817. The shaded region in the figure represents the intersection of the two constraints, offering a comprehensive view of the permissible parameter combinations for DM in NS models, based on the currently available observational data. The parameter space is defined by the two key observables:
\begin{itemize}
    \item Maximum Mass Constraint: The upper boundary (solid line) of the shaded region corresponds to the requirement that the NS maximum mass must be equal to or more than the observed mass of PSR J0740+6620, which has a lower bound of $M = 2.08^{+0.07}_{-0.07} M_{\odot}$, i.e., $M=2.01M_\odot$. Any $\alpha M_\chi$ value that yields a maximum mass above this observational threshold is excluded.
    \item Tidal Deformability Constraint: The lower boundary (dashed line) is set by the tidal deformability constraint derived from the GW170817 event. Specifically, values of $\alpha M_\chi$ that result in a tidal deformability parameter $\Lambda_{1.4}$ below the observational upper limit of 580 are allowed. This ensures that the compactness and deformability of the modeled NSs align with the behavior inferred from gravitational wave signals.
\end{itemize}

\begin{figure}[tbp]
    \centering
    \includegraphics[width=\columnwidth]{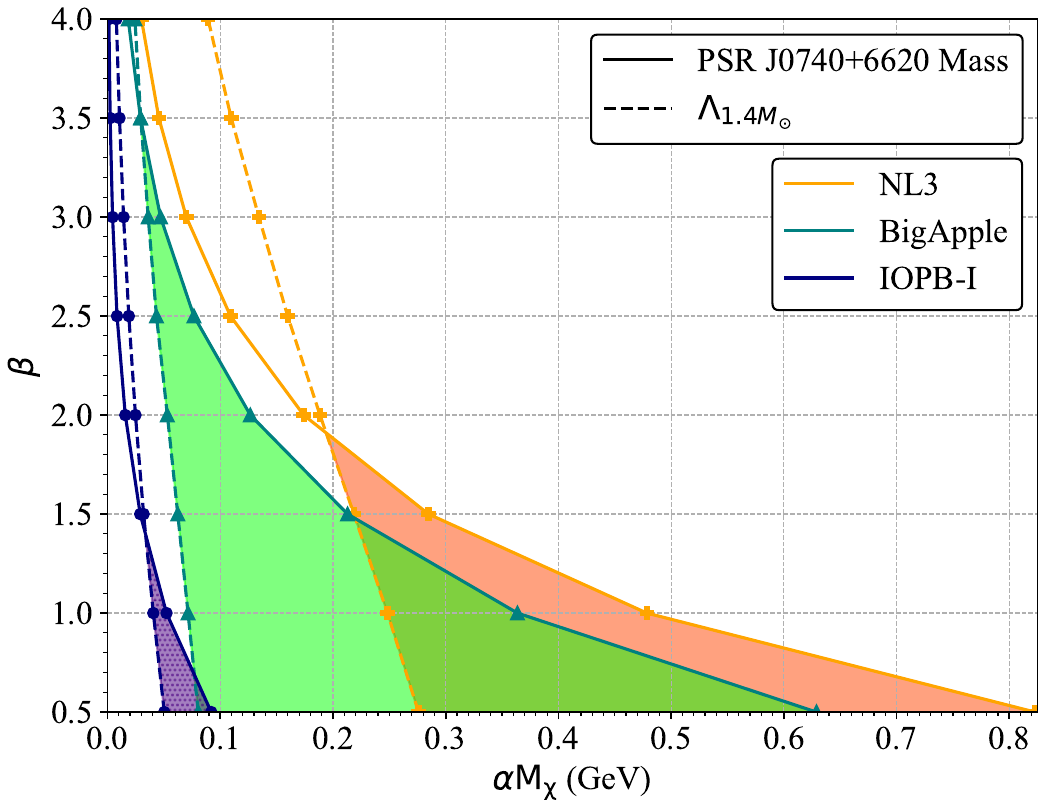}
    \caption{Same as Fig. \ref{fig:figure7}, but the allowed regions with NL3 and BigApple parameter sets are also shown.}
    \label{fig:figure8}
\end{figure}

The shaded region in Fig.~\ref{fig:figure7} represents the parameter space where both constraints are satisfied, offering a comprehensive view of how DM properties can coexist with observed NS characteristics. A clear trend emerges: as $\beta$, which dictates the steepness of the DM density profile, increases, the range of allowable $\alpha M_\chi$ values becomes more restricted. This aligns with the notion that higher $\beta$ values imply a stronger central concentration of DM, resulting in a more confined NS core. In previous analyses shown in Figs.~\ref{fig:figure4} and \ref{fig:figure5}, higher $\beta$ values led to more stringent conditions on maximum mass and tidal deformability, and this relationship is reiterated here with the narrowing of the allowed DM parameter space. The narrowing parameter space emphasizes the challenge of achieving both stability and observational compatibility as central confinement becomes more pronounced.
Figure~\ref{fig:figure8} extends the analysis of the allowed parameter space for $\beta$ versus $\alpha M_\chi$, now encompassing results from all three considered RMF parameter sets: NL3, BigApple, and IOPB-I, where the solid and dashed lines are respectively critical values obtained from the mass of PSR J0740+6620 and the tidal deformability constrained from the GW170817 event. This comprehensive visualization synthesizes insights from previous analyses, highlighting the influence of different nuclear EoS models on constraints imposed by observational data. The shaded regions in the figure represent permissible combinations of $\beta$ and $\alpha M_\chi$ for each EoS, showcasing how nuclear stiffness shapes the compatibility of DM properties with observed NS characteristics.

The variation in the allowed parameter space among the three RMF sets reflects differences in their nuclear stiffness. Stiffer EoS models, such as NL3 and BigApple, support higher maximum masses, which inherently broadens the parameter space. This outcome aligns with their ability to accommodate higher DM concentrations while maintaining stability and consistency with observational constraints from PSR J0740+6620. The broader parameter space for NL3 and BigApple indicates that they allow for greater DM admixture without exceeding observational mass limits. However, the role of tidal deformability also becomes critical in explaining the differences between NL3 and BigApple. While both are stiff EoS, BigApple exhibits lower tidal deformability across similar $\alpha M_\chi$ values compared to NL3, as observed in Fig. \ref{fig:figure5}. This lower tidal deformability implies that BigApple produces more compact NS configurations, which helps it remain within the upper limit set by GW170817 for a wider range of $\alpha M_\chi$. Conversely, the larger radii associated with NL3 result in higher tidal deformability, limiting its compatibility with $\alpha M_\chi$ values that increase compactness beyond observed constraints. This interplay between stiffness and tidal deformability highlights that the broader parameter space for BigApple arises from its ability to maintain both high maximum mass and lower tidal deformability simultaneously. In contrast, the more deformable NSs (larger value of tidal deformability) produced by NL3 face tighter constraints from the tidal deformability limit, even though its stiffness allows for higher maximum mass. Therefore, while NL3’s stiffer nature supports larger maximum masses, the resulting higher radii limit the parameter space due to the tidal deformability constraint. 

In comparison, the softer EoS (IOPB-I) results in more restricted parameter space, as it cannot sustain high maximum masses or low tidal deformability under similar DM concentrations. The constraints imposed by both observables—maximum mass and tidal deformability—become more pronounced for IOPB-I, emphasizing the significant role of nuclear matter properties in determining the compatibility of DM admixture in NSs. This comprehensive comparison of RMF sets underscores the complex but crucial interplay between DM concentration, particle mass, and nuclear interactions within NS cores. It reveals that the choice of nuclear matter EoS is not only central to modeling NS properties but also pivotal in defining the extent of DM’s compatibility with observational constraints, emphasizing the need to consider both DM and nuclear matter characteristics when modeling DM-admixed NSs.

\begin{figure*}[tbp]
    \centering
    \includegraphics[width=\textwidth]{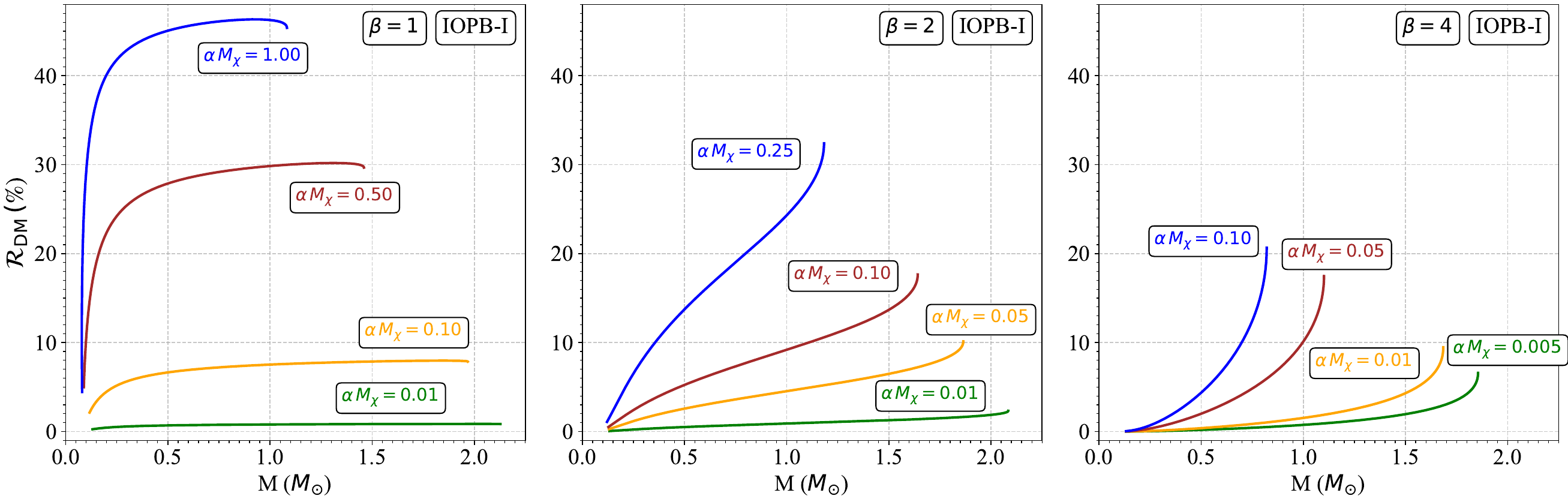}
    \caption{DM mass fraction, ${\cal R}_{\rm DM}$, as a function of NS mass, adopting the IOPB-I parameter set. The left, middle, and right panel display the results for steepness parameter $\beta =1$, 2, and 4, respectively. Each curve in a panel represents the trend for the corresponding scaling factor $\alpha M_{\chi}$.}
    \label{fig:figure9}
\end{figure*}

\subsection{DM fraction inside NSs}
\label{sec:3e}

In this subsection, we analyze the fraction of DM mass within NSs across different configurations. Specifically, we examine the proportion of gravitational mass attributed to DM, $M_{\rm DM}$, relative to the total stellar mass $M(R)$, defined as
\begin{equation}
  {\cal R}_{\rm DM} \equiv \frac{M_{\rm DM}}{M(R)}.
  \label{eq:ratio}
\end{equation}
This approach provides insights into how DM contributes to the internal structure and overall gravitational mass of NSs. The series of plots systematically investigates ${\cal R}_{\rm DM}$ based on constraints for the product $\alpha M_{\chi}$ and the steepness parameter $\beta$ derived from the observational data.

In Fig.~\ref{fig:figure9}, we examine how the DM fraction (${\cal R}_{\rm DM}$) varies with the NS mass for several configurations of $\alpha M_{\chi}$ and $\beta$ profiles, calculated with the IOPB-I RMF parameter set. Each curve has been plotted up to the maximum mass achieved by the respective $\alpha M_{\chi}$ and $\beta$ configuration, representing the highest NS mass sustainable under those DM conditions. We observe a general trend where ${\cal R}_{\rm DM}$ increases as the mass of the NS increases along each curve. This suggests that as NSs approach their mass limit, DM plays an increasingly significant role in their gravitational structure. The effect of $\beta$, which controls the steepness of the DM density profile, becomes evident when comparing ${\cal R}_{\rm DM}$ across different $\beta$ values for a fixed $\alpha M_{\chi}$. For higher $\beta$ values, such as $\beta =4$, the DM fraction for maximum mass NSs is larger, reflecting the concentrated DM distribution within the core. This enhancement in ${\cal R}_{\rm DM}$ for steeper profiles underscores DM’s role in increasing the star’s central density (see Appendix \ref{sec:appendx1}) and gravitational binding.

\begin{figure}[tbp]
    \centering
    \includegraphics[width=\columnwidth]{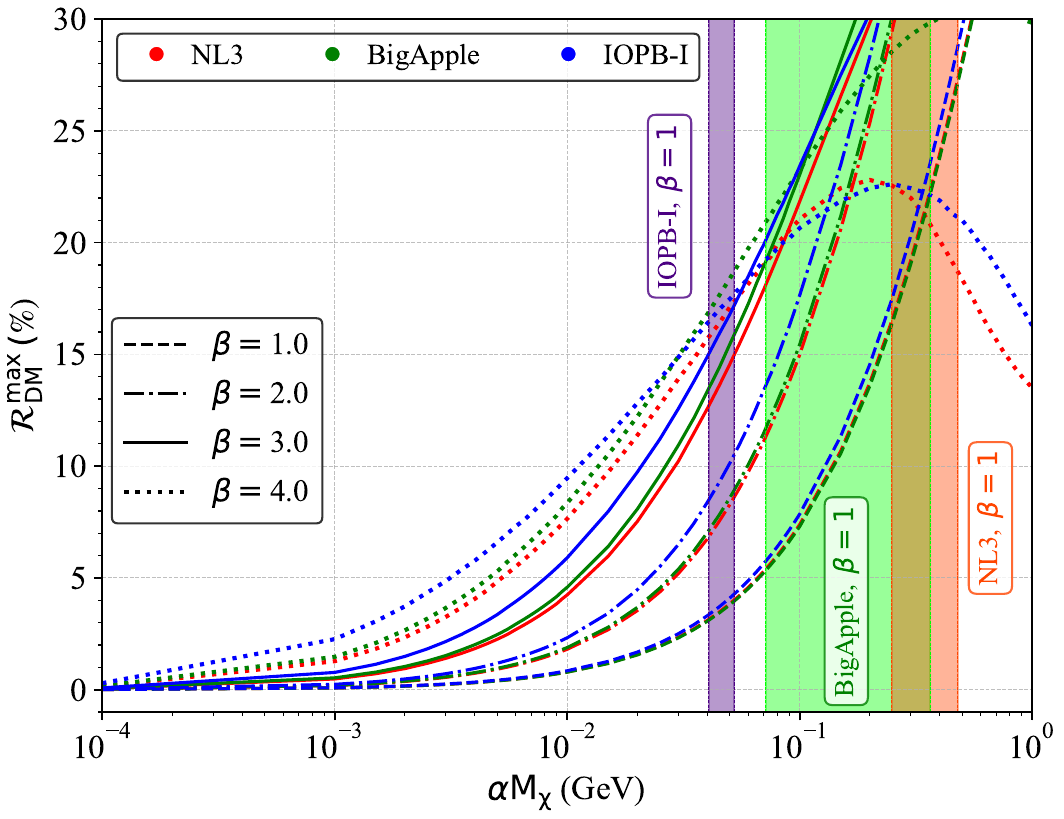}
    \caption{The DM mass fraction in the maximum mass NS for each combination of $\alpha M_{\chi}$ and $\beta$ configuration, ${\cal R}_{\rm DM}^{\rm max}$, is plotted as a function of DM concentration, $\alpha M_{\chi}$. The dashed, dash-dot-dash, solid, and dotted lines represent results for $\beta = 1$, 2, 3, and 4, respectively. Curves are color-coded for the three RMF EoSs: red for NL3, green for BigApple, and blue for IOPB-I. The shaded region indicates the observationally constrained range of $\alpha M_{\chi}$ values for $\beta = 1$, as derived in Fig. \ref{fig:figure8}, marking the parameter space consistent with the observed mass and tidal deformability constraints for PSR J0740+6620 and GW170817.}
    \label{fig:figure10}
\end{figure}

Another interesting observation in Fig. \ref{fig:figure9} is that for larger $\alpha M_{\chi}$ values, the DM fraction reaches around 20 - 40\% in low-mass NS configurations (e.g., $M\simeq 0.3M_\odot$) when $\beta = 1$, whereas this fraction is substantially lower in low-mass NSs with $\beta = 2$ or $\beta = 4$. This behavior can be attributed to the interplay between DM’s gravitational influence and its distribution across the NS, as determined by the $\alpha M_{\chi}$ and $\beta$ values. In configurations where $\beta = 1$, the DM density profile is more spread out, or less steep, resulting in a gradual DM distribution throughout the NS interior. This even distribution enables DM to occupy a larger volume within the star, thus contributing more significantly to the gravitational mass across the star, even in low-mass NSs. As a result, low-mass NSs with $\beta = 1$ and higher $\alpha M_{\chi}$ exhibit relatively high DM fractions, as DM is not strongly confined to the core and instead contributes more uniformly across the NS internal structure. Meanwhile, for higher $\beta$ values, such as $\beta = 2$ or 4, the reduced gravitational potential in low-mass NS configurations does not provide enough confinement to concentrate DM significantly within the core, which leads to a result that only a small fraction of the NS mass is attributable to DM because the steep density profile of DM, imposed by higher $\beta$, limits the volume it can occupy effectively within these lower-mass stars. Thus the DM fraction in low-mass NSs with lower $\beta$ values becomes considerably larger compared to the higher $\beta$ case. This demonstrates that the gravitational role of DM is sensitive to the steepness of its density profile. These observations emphasize that DM’s influence on NS properties extends beyond an additive contribution to the mass. Instead, it reshapes the NS’s internal mass distribution and density profile, producing configurations with distinct variations in compactness and stability that depend directly on the DM parameters.

\begin{figure}[tbp]
    \centering
    \includegraphics[width=\columnwidth]{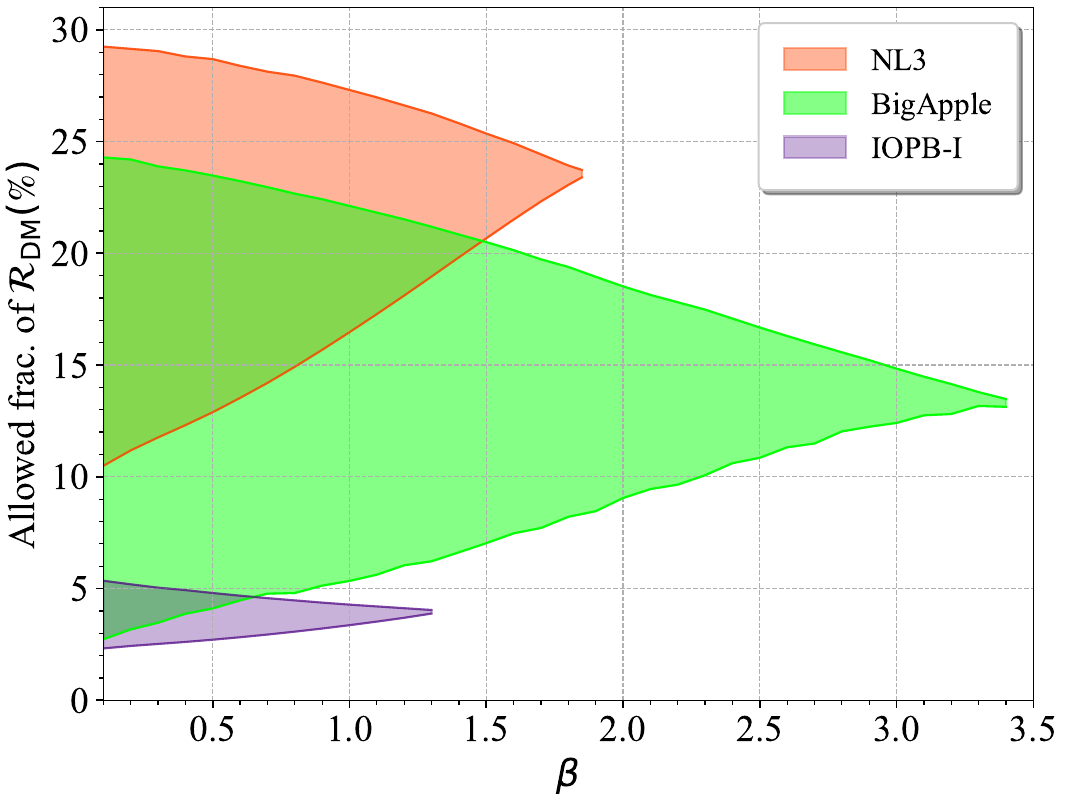}
    \caption{Allowed range of DM mass fraction, ${\cal R}_{\rm DM}$, in NSs as a function of the steepness parameter $\beta$ for each RMF parameter set considered (NL3, BigApple, and IOPB-I). The curve illustrates the maximum DM fraction permissible within each NS model under observational constraints on the mass with PSR J0740+6620 and tidal deformability with GW170817.}
    \label{fig:figure11}
\end{figure}

Following the analysis of how DM fraction ${\cal R}_{\rm DM}$ varies across NS mass configurations, Fig.~\ref{fig:figure10} depicts ${\cal R}_{\rm DM}^{\rm max}$ — the mass fraction of DM in maximum mass NSs — for different $\alpha M_{\chi}$ and $\beta$ values.  Each point represents the DM fraction within the maximum mass NS configuration for the respective $\alpha M_{\chi}$ and $\beta$, shown here for IOPB-I, BigApple, and NL3 parameter sets, where the dashed, dash-dot-dash, solid, and dotted lines represent the variation of $\beta = 1$, 2, 3, and 4, respectively. The shaded region for each parameter set corresponds to the observationally constrained range of $\alpha M_{\chi}$ with $\beta=1$ derived from Fig.~\ref{fig:figure8}. 
We note that ${\cal R}_{\rm DM}^{\rm max}$ is the value of ${\cal R}_{\rm DM}$ for the NS model with the maximum mass, but it is also corresponding to almost the maximum value of ${\cal R}_{\rm DM}$ for each parameter set, as shown in Fig.\ref{fig:figure9}).

The increase in ${\cal R}_{\rm DM}^{\rm max}$ with $\alpha M_{\chi}$ suggests that DM's contribution to the gravitational structure becomes more pronounced as the DM particle mass and scaling factor increase. Higher $\beta$ values lead to a larger DM fraction in maximum-mass NS configurations for the same $\alpha M_{\chi}$ value. This is due to the more centrally concentrated DM profile associated with higher $\beta$, which allows DM to dominate the core density and contribute more significantly to the NS's total gravitational mass. However, for $\beta = 4$, an intriguing pattern emerges at high $\alpha M_{\chi}$: ${\cal R}_{\rm DM}^{\rm max}$ initially rises but then begins to decline as $\alpha M_{\chi}$ increases further. Anyway, the range of $\alpha M_\chi$, where the value of ${\cal R}_{\rm DM}^{\rm max}$ decreases, with $\beta = 4$ has already been excluded in light of observational data because the corresponding NS mass is much smaller (see Fig.~\ref{fig:figure4}). 

In the final plot (Fig. \ref{fig:figure11}), we present the allowed range of DM fraction ${\cal R}_{\rm DM}$ as a function of $\beta$ for each RMF parameter set (IOPB-I, BigApple, and NL3). This plot consolidates the observational constraints derived in the previous figure and extends the analysis to show how $\beta$ impacts the permissible DM fraction across different nuclear matter models. In Fig. \ref{fig:figure10}, we noted that the shaded regions for each parameter set, shown specifically for $\beta = 1$, represent the allowed $\alpha M_{\chi}$ ranges based on observational constraints, which in turn define limits on the allowed DM fractions. In this figure, we generalize this approach across $\beta$ values, providing a comprehensive view of how DM fraction limits (${\cal R}_{\rm DM}$) are shaped by both the DM profile steepness and the underlying nuclear matter properties.

Distinct ranges of ${\cal R}_{\rm DM}$ appear for each RMF model, reflecting the impact of nuclear stiffness on the maximum DM fraction allowable within the NS structure. The NL3 and BigApple parameter sets, characterized by stiffer EoS, accommodate higher DM fractions compared to the softer IOPB-I model. In particular, NL3 and BigApple parameters permit relatively high DM fractions—around 29\% for NL3 and approximately 24\% for BigApple—particularly at lower $\beta$ values. This implies that NSs with stiffer nuclear matter can maintain structural stability and satisfy current observational data even with substantial DM contributions, particularly when the DM distribution is relatively uniform. These higher fractions align with the structural robustness of NSs modeled with stiffer EoS, where increased nuclear repulsion at high densities enables greater DM contributions.

Interestingly, there are regions where the permissible DM fractions of different parameter sets partially overlap at certain $\beta$ values. Specifically, we observe two types of overlaps: one where IOPB-I and BigApple align in their allowable DM fractions, and another where NL3 and BigApple overlap. This partial convergence suggests that, while each RMF model has a unique stiffness profile that affects the precise limits of DM fraction, the observational constraints on NS mass and tidal deformability enforce some structural consistency across models. This consistency ensures that even softer EoS, like IOPB-I, can accommodate moderate DM fractions under specific configurations, within limits that align with stiffer models. Thus, while stiffness directly influences the range of DM fractions, observational constraints provide a unifying framework, enforcing structural limits that align NS properties across varying EoS models. Although $\beta$ influences how DM is distributed within the NS, this figure suggests that the total permissible fraction of DM mass remains bounded primarily by the nuclear matter EoS and observational constraints, rather than by $\beta$ itself. This reinforces the notion that while DM density profiles contribute to structural variations within the NS, the fraction of DM that can be accommodated is fundamentally limited by the underlying nuclear matter properties and astrophysical observations. Overall, this result underscores the study’s broader conclusions, integrating DM parameters, nuclear EoS, and observational constraints into a cohesive framework for understanding DM’s influence on neutron star properties.

\section{Conclusion}
\label{sec:4}

This study offers an in-depth exploration of the effects of DM distribution on NS properties, drawing from a range of relativistic mean-field models and incorporating the current observational constraints. Through a systematic examination of density profiles, mass-radius relations, maximum mass limits, tidal deformability, and DM fractions, we have elucidated how DM parameters, nuclear matter EoS, and observational benchmarks collectively shape the structural characteristics of NSs. Key findings from this work provide a cohesive framework that integrates DM’s role within the dense environments of NS interiors and highlights its potential detectability through gravitational wave and pulsar observations. The DM’s contribution to energy density, rather than pressure, underlines its unique role in modifying NS structure. Unlike baryonic matter, which contributes to energy density and pressure, DM predominantly enhances the gravitational field without significantly altering the pressure balance. This distinct behavior highlights the need for models incorporating DM contributions primarily through energy density to accurately describe the core structure. 

The study reveals that DM influences NS structure primarily through its contribution to gravitational confinement, with the product $\alpha M_{\chi}$ dictating the degree of compactness. The analysis shows that NSs with higher $\alpha M_{\chi}$ values exhibit increased compactness, leading to reduced radii and tidal deformability, particularly for steep DM profiles (higher $\beta$). The mass-radius profiles for constant $\alpha M_{\chi}$ emphasize that DM energy density is the primary factor governing NS compactness rather than individual variations in $\alpha$ or $M_{\chi}$ and supports the notion that gravitational confinement is driven by total DM density rather than specific distribution characteristics alone.

Importantly, our analysis demonstrates that the maximum mass and tidal deformability of DM-admixed NSs are sensitive to nuclear stiffness, with stiffer RMF models such as NL3 and BigApple supporting higher DM fractions compared to softer models like IOPB-I. Observational constraints, notably from PSR J0740+6620 and GW170817, impose stringent limits on the DM parameter space, indicating that high central DM concentration (i.e., higher $\beta$ values) leads to configurations that may become overly compact, conflicting with observed NS properties. Interestingly, the study finds partial convergence in the permissible DM fractions across models, despite their different nuclear stiffness, underscoring that observational limits act as unifying constraints on DM fraction, creating structural compatibility across diverse EoS models.

In examining DM fraction limits (${\cal R}_{\rm DM}$), we find that stiffer EoS models allow for larger permissible DM fractions, particularly for more uniform DM distributions. However, the observational constraints significantly bound these values, reinforcing that NS stability under DM influence is fundamentally governed by the nuclear EoS and astrophysical measurements. This finding highlights a universal structural limit on DM’s contribution across NS models, suggesting that the allowed DM fraction is not solely determined by DM density profile steepness ($\beta$) but is also shaped by nuclear matter properties and astrophysical constraints. 

The increased compactness of NSs due to DM admixture, particularly for higher $\alpha M_\chi$ and $\beta$, may influence gravitational wave signals from NS mergers by altering tidal deformability. This potential modification could serve as an observable signature in future gravitational wave detections, providing a new avenue for probing the presence of DM within NS interiors. Alternative theories of gravity could also lead to similar observations in tidal deformability. However, distinguishing the effects of DM from those caused by alternative theories of gravity remains challenging with current observational precision. A more detailed exploration of these differences, while beyond the scope of this work, could provide valuable insights and warrants further investigation. While the increased DM concentration results in more compact NS configurations, the stability of such dense cores remains an important consideration. Higher $\alpha M_\chi$ and $\beta$ values suggest central DM concentrations that could raise questions about potential instabilities. The exploration of stability criteria for DM-admixed NSs, particularly under extreme central DM densities, represents an essential avenue for future research. Future studies may extend this work by exploring DM density profiles that evolve dynamically in response to NS environments, considering both core and crustal effects. Another phenomenon worth considering is the annihilation of DM particles, which may occur in the DM model explored in this study. While its effects were not explicitly included in our analysis, such annihilation could introduce additional mechanisms influencing NS properties, such as mass, radius, and thermal signatures, representing a valuable extension to the current model. Additionally, as gravitational wave detections improve, refined tidal deformability measurements could provide further constraints on DM presence in NSs, offering deeper insights into the interplay between DM and nuclear matter in ultra-dense astrophysical objects.

\acknowledgments

This work is supported in part by Japan Society for the Promotion of Science (JSPS) KAKENHI Grant Numbers 
JP23K20848  
and JP24KF0090. 

\appendix
\section{Mass dependence of the central energy density}
\label{sec:appendx1}

\begin{figure*}[tbp]
    \centering
    \includegraphics[width=\textwidth]{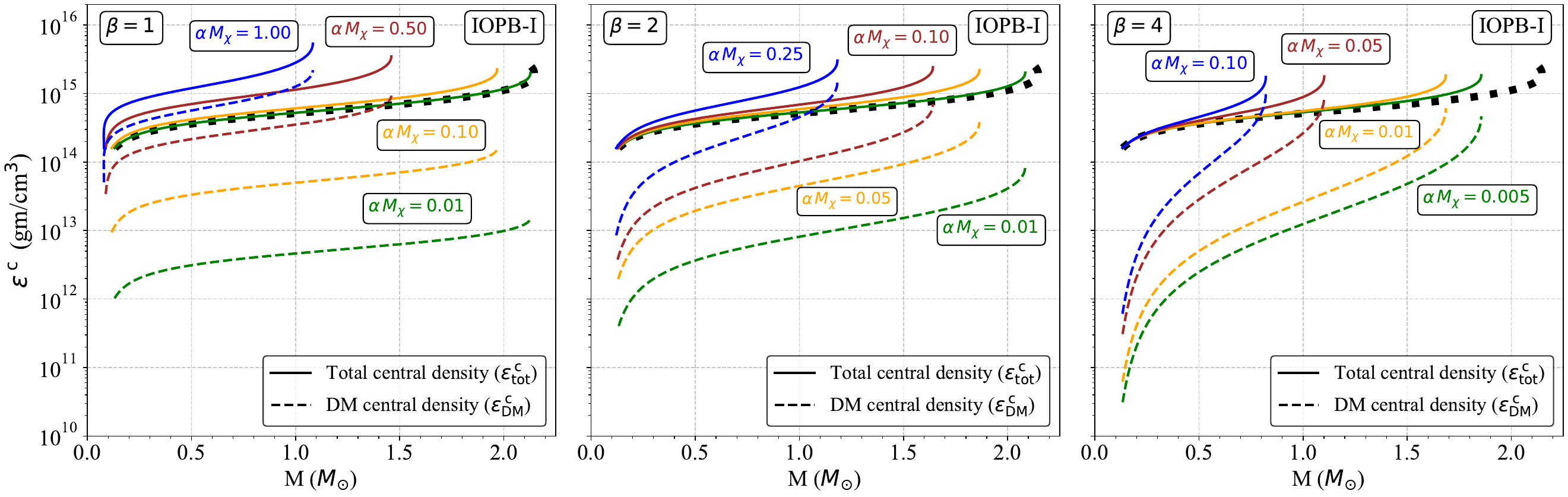}
    \caption{The NS mass dependence of the central energy density for a total of DM and baryonic matter, $\varepsilon_{\rm tot}^{c}$, and that for DM, $\varepsilon_{\rm DM}^c$, are respectively shown with the solid and dashed lines, adopting the IOPB-I parameter set for various values of $\alpha M_\chi$. The left, middle, and right panels correspond to the results with $\beta =1$, 2, and 4. For reference, the dotted line denotes the central energy density for NS models constructed without DM.}
    \label{fig:figure12}
\end{figure*}

Figure \ref{fig:figure12} presents the total central energy density ($\varepsilon_{\rm tot}^{\rm c}$) and DM central energy density ($\varepsilon_{\rm DM}^{\rm c}$) for various NS models, using the IOPB-I parameter set. The solid lines indicate $\varepsilon_{\rm tot}^{\rm \,c}$, while the dashed lines represent $\varepsilon_{\rm DM}^{\rm \,c}$. This figure highlights how central energy densities change as $\alpha M_\chi$ varies, illustrating the growing influence of DM in the NS core. As $\beta$ increases, the central density contributed by DM approaches the total central density of the NS, intensifying the gravitational pull within the core region. This elevated central density heightens the risk of instability under further DM influence, effectively limiting the DM fraction that can be sustained without exceeding the NS’s stability.

As the NS mass increases, both $\varepsilon_{\rm tot}^{\rm c}$ and $\varepsilon_{\rm DM}^{\rm c}$ rise, reflecting the stronger gravitational attraction induced by DM concentration. However, the most critical insight from this plot is the potential constraint on DM properties through NS stability. As $\varepsilon_{\rm DM}^{\rm c}$ becomes a larger fraction of $\varepsilon_{\rm tot}^{\rm c}$, the baryonic pressure must increase to counter the added gravitational attraction from DM. If $\varepsilon_{\rm DM}^{\rm c}$ reaches a level where the baryonic pressure can no longer provide sufficient support, the NS becomes unstable and collapses.

This threshold for stability represents a critical limit for DM admixture in NSs. By identifying the maximum $\varepsilon_{\rm DM}^{\rm c}$ that still permits a stable NS configuration, it becomes possible to set upper bounds on $\alpha M_\chi$, and thereby indirectly constrain DM concentration or particle mass. These constraints are linked to the observational limits, such as those set by PSR J0740+6620, which sets the stability threshold for NSs with DM admixture.

While the current study does not explore this threshold in detail, it offers an avenue for future work. By conducting more detailed investigations into the critical $\varepsilon_{\rm DM}^{\rm \,c}$ required for NS collapse, it could be possible to establish even tighter constraints on DM properties. Such an approach may extend beyond compact objects, potentially shedding light on DM characteristics in broader astrophysical contexts, such as galactic halos.

\bibliographystyle{apsrev4-2}
\bibliography{main.bib} 

\begin{thebibliography}{64}%
\makeatletter
\providecommand \@ifxundefined [1]{%
 \@ifx{#1\undefined}
}%
\providecommand \@ifnum [1]{%
 \ifnum #1\expandafter \@firstoftwo
 \else \expandafter \@secondoftwo
 \fi
}%
\providecommand \@ifx [1]{%
 \ifx #1\expandafter \@firstoftwo
 \else \expandafter \@secondoftwo
 \fi
}%
\providecommand \natexlab [1]{#1}%
\providecommand \enquote  [1]{``#1''}%
\providecommand \bibnamefont  [1]{#1}%
\providecommand \bibfnamefont [1]{#1}%
\providecommand \citenamefont [1]{#1}%
\providecommand \href@noop [0]{\@secondoftwo}%
\providecommand \href [0]{\begingroup \@sanitize@url \@href}%
\providecommand \@href[1]{\@@startlink{#1}\@@href}%
\providecommand \@@href[1]{\endgroup#1\@@endlink}%
\providecommand \@sanitize@url [0]{\catcode `\\12\catcode `\$12\catcode `\&12\catcode `\#12\catcode `\^12\catcode `\_12\catcode `\%12\relax}%
\providecommand \@@startlink[1]{}%
\providecommand \@@endlink[0]{}%
\providecommand \url  [0]{\begingroup\@sanitize@url \@url }%
\providecommand \@url [1]{\endgroup\@href {#1}{\urlprefix }}%
\providecommand \urlprefix  [0]{URL }%
\providecommand \Eprint [0]{\href }%
\providecommand \doibase [0]{https://doi.org/}%
\providecommand \selectlanguage [0]{\@gobble}%
\providecommand \bibinfo  [0]{\@secondoftwo}%
\providecommand \bibfield  [0]{\@secondoftwo}%
\providecommand \translation [1]{[#1]}%
\providecommand \BibitemOpen [0]{}%
\providecommand \bibitemStop [0]{}%
\providecommand \bibitemNoStop [0]{.\EOS\space}%
\providecommand \EOS [0]{\spacefactor3000\relax}%
\providecommand \BibitemShut  [1]{\csname bibitem#1\endcsname}%
\let\auto@bib@innerbib\@empty
\bibitem [{\citenamefont {Alford}\ \emph {et~al.}(2008)\citenamefont {Alford}, \citenamefont {Schmitt}, \citenamefont {Rajagopal},\ and\ \citenamefont {Sch\"afer}}]{RevModPhys.80.1455}%
  \BibitemOpen
  \bibfield  {author} {\bibinfo {author} {\bibfnamefont {M.~G.}\ \bibnamefont {Alford}}, \bibinfo {author} {\bibfnamefont {A.}~\bibnamefont {Schmitt}}, \bibinfo {author} {\bibfnamefont {K.}~\bibnamefont {Rajagopal}},\ and\ \bibinfo {author} {\bibfnamefont {T.}~\bibnamefont {Sch\"afer}},\ }\href {https://doi.org/10.1103/RevModPhys.80.1455} {\bibfield  {journal} {\bibinfo  {journal} {Rev. Mod. Phys.}\ }\textbf {\bibinfo {volume} {80}},\ \bibinfo {pages} {1455} (\bibinfo {year} {2008})}\BibitemShut {NoStop}%
\bibitem [{\citenamefont {Fattoyev}\ \emph {et~al.}(2017)\citenamefont {Fattoyev}, \citenamefont {Horowitz},\ and\ \citenamefont {Schuetrumpf}}]{PhysRevC.95.055804}%
  \BibitemOpen
  \bibfield  {author} {\bibinfo {author} {\bibfnamefont {F.~J.}\ \bibnamefont {Fattoyev}}, \bibinfo {author} {\bibfnamefont {C.~J.}\ \bibnamefont {Horowitz}},\ and\ \bibinfo {author} {\bibfnamefont {B.}~\bibnamefont {Schuetrumpf}},\ }\href {https://doi.org/10.1103/PhysRevC.95.055804} {\bibfield  {journal} {\bibinfo  {journal} {Phys. Rev. C}\ }\textbf {\bibinfo {volume} {95}},\ \bibinfo {pages} {055804} (\bibinfo {year} {2017})}\BibitemShut {NoStop}%
\bibitem [{\citenamefont {Vidaña}\ \emph {et~al.}(2018)\citenamefont {Vidaña}, \citenamefont {Bashkanov}, \citenamefont {Watts},\ and\ \citenamefont {Pastore}}]{VIDANA2018112}%
  \BibitemOpen
  \bibfield  {author} {\bibinfo {author} {\bibfnamefont {I.}~\bibnamefont {Vidaña}}, \bibinfo {author} {\bibfnamefont {M.}~\bibnamefont {Bashkanov}}, \bibinfo {author} {\bibfnamefont {D.}~\bibnamefont {Watts}},\ and\ \bibinfo {author} {\bibfnamefont {A.}~\bibnamefont {Pastore}},\ }\href {https://doi.org/https://doi.org/10.1016/j.physletb.2018.03.052} {\bibfield  {journal} {\bibinfo  {journal} {Physics Letters B}\ }\textbf {\bibinfo {volume} {781}},\ \bibinfo {pages} {112} (\bibinfo {year} {2018})}\BibitemShut {NoStop}%
\bibitem [{\citenamefont {Tolman}(1939)}]{PhysRev.55.364}%
  \BibitemOpen
  \bibfield  {author} {\bibinfo {author} {\bibfnamefont {R.~C.}\ \bibnamefont {Tolman}},\ }\href {https://doi.org/10.1103/PhysRev.55.364} {\bibfield  {journal} {\bibinfo  {journal} {Phys. Rev.}\ }\textbf {\bibinfo {volume} {55}},\ \bibinfo {pages} {364} (\bibinfo {year} {1939})}\BibitemShut {NoStop}%
\bibitem [{\citenamefont {Oppenheimer}\ and\ \citenamefont {Volkoff}(1939)}]{PhysRev.55.374}%
  \BibitemOpen
  \bibfield  {author} {\bibinfo {author} {\bibfnamefont {J.~R.}\ \bibnamefont {Oppenheimer}}\ and\ \bibinfo {author} {\bibfnamefont {G.~M.}\ \bibnamefont {Volkoff}},\ }\href {https://doi.org/10.1103/PhysRev.55.374} {\bibfield  {journal} {\bibinfo  {journal} {Phys. Rev.}\ }\textbf {\bibinfo {volume} {55}},\ \bibinfo {pages} {374} (\bibinfo {year} {1939})}\BibitemShut {NoStop}%
\bibitem [{\citenamefont {Bogdanov}\ \emph {et~al.}(2019)\citenamefont {Bogdanov}, \citenamefont {Lamb}, \citenamefont {Mahmoodifar}, \citenamefont {Miller}, \citenamefont {Morsink}, \citenamefont {Riley}, \citenamefont {Strohmayer}, \citenamefont {Tung}, \citenamefont {Watts}, \citenamefont {Dittmann}, \citenamefont {Chakrabarty}, \citenamefont {Guillot}, \citenamefont {Arzoumanian},\ and\ \citenamefont {Gendreau}}]{Bogdanov_2019}%
  \BibitemOpen
  \bibfield  {author} {\bibinfo {author} {\bibfnamefont {S.}~\bibnamefont {Bogdanov}}, \bibinfo {author} {\bibfnamefont {F.~K.}\ \bibnamefont {Lamb}}, \bibinfo {author} {\bibfnamefont {S.}~\bibnamefont {Mahmoodifar}}, \bibinfo {author} {\bibfnamefont {M.~C.}\ \bibnamefont {Miller}}, \bibinfo {author} {\bibfnamefont {S.~M.}\ \bibnamefont {Morsink}}, \bibinfo {author} {\bibfnamefont {T.~E.}\ \bibnamefont {Riley}}, \bibinfo {author} {\bibfnamefont {T.~E.}\ \bibnamefont {Strohmayer}}, \bibinfo {author} {\bibfnamefont {A.~K.}\ \bibnamefont {Tung}}, \bibinfo {author} {\bibfnamefont {A.~L.}\ \bibnamefont {Watts}}, \bibinfo {author} {\bibfnamefont {A.~J.}\ \bibnamefont {Dittmann}}, \bibinfo {author} {\bibfnamefont {D.}~\bibnamefont {Chakrabarty}}, \bibinfo {author} {\bibfnamefont {S.}~\bibnamefont {Guillot}}, \bibinfo {author} {\bibfnamefont {Z.}~\bibnamefont {Arzoumanian}},\ and\ \bibinfo {author} {\bibfnamefont {K.~C.}\ \bibnamefont {Gendreau}},\ }\href {https://doi.org/10.3847/2041-8213/ab5968} {\bibfield
  {journal} {\bibinfo  {journal} {The Astrophysical Journal Letters}\ }\textbf {\bibinfo {volume} {887}},\ \bibinfo {pages} {L26} (\bibinfo {year} {2019})}\BibitemShut {NoStop}%
\bibitem [{\citenamefont {Bogdanov}\ \emph {et~al.}(2021)\citenamefont {Bogdanov}, \citenamefont {Dittmann}, \citenamefont {Ho}, \citenamefont {Lamb}, \citenamefont {Mahmoodifar}, \citenamefont {Miller}, \citenamefont {Morsink}, \citenamefont {Riley}, \citenamefont {Strohmayer}, \citenamefont {Watts}, \citenamefont {Choudhury}, \citenamefont {Guillot}, \citenamefont {Harding}, \citenamefont {Ray}, \citenamefont {Wadiasingh}, \citenamefont {Wolff}, \citenamefont {Markwardt}, \citenamefont {Arzoumanian},\ and\ \citenamefont {Gendreau}}]{Bogdanov_2021}%
  \BibitemOpen
  \bibfield  {author} {\bibinfo {author} {\bibfnamefont {S.}~\bibnamefont {Bogdanov}}, \bibinfo {author} {\bibfnamefont {A.~J.}\ \bibnamefont {Dittmann}}, \bibinfo {author} {\bibfnamefont {W.~C.~G.}\ \bibnamefont {Ho}}, \bibinfo {author} {\bibfnamefont {F.~K.}\ \bibnamefont {Lamb}}, \bibinfo {author} {\bibfnamefont {S.}~\bibnamefont {Mahmoodifar}}, \bibinfo {author} {\bibfnamefont {M.~C.}\ \bibnamefont {Miller}}, \bibinfo {author} {\bibfnamefont {S.~M.}\ \bibnamefont {Morsink}}, \bibinfo {author} {\bibfnamefont {T.~E.}\ \bibnamefont {Riley}}, \bibinfo {author} {\bibfnamefont {T.~E.}\ \bibnamefont {Strohmayer}}, \bibinfo {author} {\bibfnamefont {A.~L.}\ \bibnamefont {Watts}}, \bibinfo {author} {\bibfnamefont {D.}~\bibnamefont {Choudhury}}, \bibinfo {author} {\bibfnamefont {S.}~\bibnamefont {Guillot}}, \bibinfo {author} {\bibfnamefont {A.~K.}\ \bibnamefont {Harding}}, \bibinfo {author} {\bibfnamefont {P.~S.}\ \bibnamefont {Ray}}, \bibinfo {author} {\bibfnamefont {Z.}~\bibnamefont {Wadiasingh}}, \bibinfo
  {author} {\bibfnamefont {M.~T.}\ \bibnamefont {Wolff}}, \bibinfo {author} {\bibfnamefont {C.~B.}\ \bibnamefont {Markwardt}}, \bibinfo {author} {\bibfnamefont {Z.}~\bibnamefont {Arzoumanian}},\ and\ \bibinfo {author} {\bibfnamefont {K.~C.}\ \bibnamefont {Gendreau}},\ }\href {https://doi.org/10.3847/2041-8213/abfb79} {\bibfield  {journal} {\bibinfo  {journal} {The Astrophysical Journal Letters}\ }\textbf {\bibinfo {volume} {914}},\ \bibinfo {pages} {L15} (\bibinfo {year} {2021})}\BibitemShut {NoStop}%
\bibitem [{\citenamefont {Jarosik}\ \emph {et~al.}(2011)\citenamefont {Jarosik}, \citenamefont {Bennett}, \citenamefont {Dunkley}, \citenamefont {Gold}, \citenamefont {Greason}, \citenamefont {Halpern}, \citenamefont {Hill}, \citenamefont {Hinshaw}, \citenamefont {Kogut}, \citenamefont {Komatsu}, \citenamefont {Larson}, \citenamefont {Limon}, \citenamefont {Meyer}, \citenamefont {Nolta}, \citenamefont {Odegard}, \citenamefont {Page}, \citenamefont {Smith}, \citenamefont {Spergel}, \citenamefont {Tucker}, \citenamefont {Weiland}, \citenamefont {Wollack},\ and\ \citenamefont {Wright}}]{Jarosik_2011}%
  \BibitemOpen
  \bibfield  {author} {\bibinfo {author} {\bibfnamefont {N.}~\bibnamefont {Jarosik}}, \bibinfo {author} {\bibfnamefont {C.~L.}\ \bibnamefont {Bennett}}, \bibinfo {author} {\bibfnamefont {J.}~\bibnamefont {Dunkley}}, \bibinfo {author} {\bibfnamefont {B.}~\bibnamefont {Gold}}, \bibinfo {author} {\bibfnamefont {M.~R.}\ \bibnamefont {Greason}}, \bibinfo {author} {\bibfnamefont {M.}~\bibnamefont {Halpern}}, \bibinfo {author} {\bibfnamefont {R.~S.}\ \bibnamefont {Hill}}, \bibinfo {author} {\bibfnamefont {G.}~\bibnamefont {Hinshaw}}, \bibinfo {author} {\bibfnamefont {A.}~\bibnamefont {Kogut}}, \bibinfo {author} {\bibfnamefont {E.}~\bibnamefont {Komatsu}}, \bibinfo {author} {\bibfnamefont {D.}~\bibnamefont {Larson}}, \bibinfo {author} {\bibfnamefont {M.}~\bibnamefont {Limon}}, \bibinfo {author} {\bibfnamefont {S.~S.}\ \bibnamefont {Meyer}}, \bibinfo {author} {\bibfnamefont {M.~R.}\ \bibnamefont {Nolta}}, \bibinfo {author} {\bibfnamefont {N.}~\bibnamefont {Odegard}}, \bibinfo {author} {\bibfnamefont {L.}~\bibnamefont
  {Page}}, \bibinfo {author} {\bibfnamefont {K.~M.}\ \bibnamefont {Smith}}, \bibinfo {author} {\bibfnamefont {D.~N.}\ \bibnamefont {Spergel}}, \bibinfo {author} {\bibfnamefont {G.~S.}\ \bibnamefont {Tucker}}, \bibinfo {author} {\bibfnamefont {J.~L.}\ \bibnamefont {Weiland}}, \bibinfo {author} {\bibfnamefont {E.}~\bibnamefont {Wollack}},\ and\ \bibinfo {author} {\bibfnamefont {E.~L.}\ \bibnamefont {Wright}},\ }\href {https://doi.org/10.1088/0067-0049/192/2/14} {\bibfield  {journal} {\bibinfo  {journal} {The Astrophysical Journal Supplement Series}\ }\textbf {\bibinfo {volume} {192}},\ \bibinfo {pages} {14} (\bibinfo {year} {2011})}\BibitemShut {NoStop}%
\bibitem [{\citenamefont {Bertone}\ \emph {et~al.}(2005)\citenamefont {Bertone}, \citenamefont {Hooper},\ and\ \citenamefont {Silk}}]{BERTONE2005279}%
  \BibitemOpen
  \bibfield  {author} {\bibinfo {author} {\bibfnamefont {G.}~\bibnamefont {Bertone}}, \bibinfo {author} {\bibfnamefont {D.}~\bibnamefont {Hooper}},\ and\ \bibinfo {author} {\bibfnamefont {J.}~\bibnamefont {Silk}},\ }\href {https://doi.org/https://doi.org/10.1016/j.physrep.2004.08.031} {\bibfield  {journal} {\bibinfo  {journal} {Physics Reports}\ }\textbf {\bibinfo {volume} {405}},\ \bibinfo {pages} {279} (\bibinfo {year} {2005})}\BibitemShut {NoStop}%
\bibitem [{\citenamefont {Goldman}\ and\ \citenamefont {Nussinov}(1989)}]{PhysRevD.40.3221}%
  \BibitemOpen
  \bibfield  {author} {\bibinfo {author} {\bibfnamefont {I.}~\bibnamefont {Goldman}}\ and\ \bibinfo {author} {\bibfnamefont {S.}~\bibnamefont {Nussinov}},\ }\href {https://doi.org/10.1103/PhysRevD.40.3221} {\bibfield  {journal} {\bibinfo  {journal} {Phys. Rev. D}\ }\textbf {\bibinfo {volume} {40}},\ \bibinfo {pages} {3221} (\bibinfo {year} {1989})}\BibitemShut {NoStop}%
\bibitem [{\citenamefont {Ray}\ \emph {et~al.}(2019)\citenamefont {Ray}, \citenamefont {Arzoumanian}, \citenamefont {Ballantyne}, \citenamefont {Bozzo}, \citenamefont {Brandt}, \citenamefont {Brenneman} \emph {et~al.}}]{ray2019strobexxraytimingspectroscopy}%
  \BibitemOpen
  \bibfield  {author} {\bibinfo {author} {\bibfnamefont {P.~S.}\ \bibnamefont {Ray}}, \bibinfo {author} {\bibfnamefont {Z.}~\bibnamefont {Arzoumanian}}, \bibinfo {author} {\bibfnamefont {D.}~\bibnamefont {Ballantyne}}, \bibinfo {author} {\bibfnamefont {E.}~\bibnamefont {Bozzo}}, \bibinfo {author} {\bibfnamefont {S.}~\bibnamefont {Brandt}}, \bibinfo {author} {\bibfnamefont {L.}~\bibnamefont {Brenneman}}, \emph {et~al.},\ }\href {https://arxiv.org/abs/1903.03035} {\bibinfo {title} {Strobe-x: X-ray timing and spectroscopy on dynamical timescales from microseconds to years}} (\bibinfo {year} {2019}),\ \Eprint {https://arxiv.org/abs/1903.03035} {arXiv:1903.03035 [astro-ph.IM]} \BibitemShut {NoStop}%
\bibitem [{\citenamefont {Watts}\ \emph {et~al.}(2018)\citenamefont {Watts}, \citenamefont {Yu}, \citenamefont {Poutanen}, \citenamefont {Zhang}, \citenamefont {Bhattacharyya} \emph {et~al.}}]{Watts2018}%
  \BibitemOpen
  \bibfield  {author} {\bibinfo {author} {\bibfnamefont {A.~L.}\ \bibnamefont {Watts}}, \bibinfo {author} {\bibfnamefont {W.}~\bibnamefont {Yu}}, \bibinfo {author} {\bibfnamefont {J.}~\bibnamefont {Poutanen}}, \bibinfo {author} {\bibfnamefont {S.}~\bibnamefont {Zhang}}, \bibinfo {author} {\bibfnamefont {S.}~\bibnamefont {Bhattacharyya}}, \emph {et~al.},\ }\href {https://doi.org/10.1007/s11433-017-9188-4} {\bibfield  {journal} {\bibinfo  {journal} {Science China Physics, Mechanics {\&} Astronomy}\ }\textbf {\bibinfo {volume} {62}},\ \bibinfo {pages} {29503} (\bibinfo {year} {2018})}\BibitemShut {NoStop}%
\bibitem [{\citenamefont {Kaup}(1968)}]{PhysRev.172.1331}%
  \BibitemOpen
  \bibfield  {author} {\bibinfo {author} {\bibfnamefont {D.~J.}\ \bibnamefont {Kaup}},\ }\href {https://doi.org/10.1103/PhysRev.172.1331} {\bibfield  {journal} {\bibinfo  {journal} {Phys. Rev.}\ }\textbf {\bibinfo {volume} {172}},\ \bibinfo {pages} {1331} (\bibinfo {year} {1968})}\BibitemShut {NoStop}%
\bibitem [{\citenamefont {Colpi}\ \emph {et~al.}(1986)\citenamefont {Colpi}, \citenamefont {Shapiro},\ and\ \citenamefont {Wasserman}}]{PhysRevLett.57.2485}%
  \BibitemOpen
  \bibfield  {author} {\bibinfo {author} {\bibfnamefont {M.}~\bibnamefont {Colpi}}, \bibinfo {author} {\bibfnamefont {S.~L.}\ \bibnamefont {Shapiro}},\ and\ \bibinfo {author} {\bibfnamefont {I.}~\bibnamefont {Wasserman}},\ }\href {https://doi.org/10.1103/PhysRevLett.57.2485} {\bibfield  {journal} {\bibinfo  {journal} {Phys. Rev. Lett.}\ }\textbf {\bibinfo {volume} {57}},\ \bibinfo {pages} {2485} (\bibinfo {year} {1986})}\BibitemShut {NoStop}%
\bibitem [{\citenamefont {Das}\ \emph {et~al.}(2022)\citenamefont {Das}, \citenamefont {Malik},\ and\ \citenamefont {Nayak}}]{PhysRevD.105.123034}%
  \BibitemOpen
  \bibfield  {author} {\bibinfo {author} {\bibfnamefont {A.}~\bibnamefont {Das}}, \bibinfo {author} {\bibfnamefont {T.}~\bibnamefont {Malik}},\ and\ \bibinfo {author} {\bibfnamefont {A.~C.}\ \bibnamefont {Nayak}},\ }\href {https://doi.org/10.1103/PhysRevD.105.123034} {\bibfield  {journal} {\bibinfo  {journal} {Phys. Rev. D}\ }\textbf {\bibinfo {volume} {105}},\ \bibinfo {pages} {123034} (\bibinfo {year} {2022})}\BibitemShut {NoStop}%
\bibitem [{\citenamefont {Vikiaris}(2024)}]{particles7030040}%
  \BibitemOpen
  \bibfield  {author} {\bibinfo {author} {\bibfnamefont {M.}~\bibnamefont {Vikiaris}},\ }\href {https://doi.org/10.3390/particles7030040} {\bibfield  {journal} {\bibinfo  {journal} {Particles}\ }\textbf {\bibinfo {volume} {7}},\ \bibinfo {pages} {692} (\bibinfo {year} {2024})}\BibitemShut {NoStop}%
\bibitem [{\citenamefont {Ivanytskyi}\ \emph {et~al.}(2020)\citenamefont {Ivanytskyi}, \citenamefont {Sagun},\ and\ \citenamefont {Lopes}}]{PhysRevD.102.063028}%
  \BibitemOpen
  \bibfield  {author} {\bibinfo {author} {\bibfnamefont {O.}~\bibnamefont {Ivanytskyi}}, \bibinfo {author} {\bibfnamefont {V.}~\bibnamefont {Sagun}},\ and\ \bibinfo {author} {\bibfnamefont {I.}~\bibnamefont {Lopes}},\ }\href {https://doi.org/10.1103/PhysRevD.102.063028} {\bibfield  {journal} {\bibinfo  {journal} {Phys. Rev. D}\ }\textbf {\bibinfo {volume} {102}},\ \bibinfo {pages} {063028} (\bibinfo {year} {2020})}\BibitemShut {NoStop}%
\bibitem [{\citenamefont {Mariani}\ \emph {et~al.}(2023)\citenamefont {Mariani}, \citenamefont {Albertus}, \citenamefont {Alessandroni}, \citenamefont {Orsaria}, \citenamefont {Perez-Garcia},\ and\ \citenamefont {Ranea-Sandoval}}]{10.1093mnrasstad3658}%
  \BibitemOpen
  \bibfield  {author} {\bibinfo {author} {\bibfnamefont {M.}~\bibnamefont {Mariani}}, \bibinfo {author} {\bibfnamefont {C.}~\bibnamefont {Albertus}}, \bibinfo {author} {\bibfnamefont {M.~d.~R.}\ \bibnamefont {Alessandroni}}, \bibinfo {author} {\bibfnamefont {M.~G.}\ \bibnamefont {Orsaria}}, \bibinfo {author} {\bibfnamefont {M.~A.}\ \bibnamefont {Perez-Garcia}},\ and\ \bibinfo {author} {\bibfnamefont {I.~F.}\ \bibnamefont {Ranea-Sandoval}},\ }\href {https://doi.org/10.1093/mnras/stad3658} {\bibfield  {journal} {\bibinfo  {journal} {Monthly Notices of the Royal Astronomical Society}\ }\textbf {\bibinfo {volume} {527}},\ \bibinfo {pages} {6795} (\bibinfo {year} {2023})}\BibitemShut {NoStop}%
\bibitem [{\citenamefont {Das}\ \emph {et~al.}(2021{\natexlab{a}})\citenamefont {Das}, \citenamefont {Kumar},\ and\ \citenamefont {Patra}}]{PhysRevD.104.063028}%
  \BibitemOpen
  \bibfield  {author} {\bibinfo {author} {\bibfnamefont {H.~C.}\ \bibnamefont {Das}}, \bibinfo {author} {\bibfnamefont {A.}~\bibnamefont {Kumar}},\ and\ \bibinfo {author} {\bibfnamefont {S.~K.}\ \bibnamefont {Patra}},\ }\href {https://doi.org/10.1103/PhysRevD.104.063028} {\bibfield  {journal} {\bibinfo  {journal} {Phys. Rev. D}\ }\textbf {\bibinfo {volume} {104}},\ \bibinfo {pages} {063028} (\bibinfo {year} {2021}{\natexlab{a}})}\BibitemShut {NoStop}%
\bibitem [{\citenamefont {Ciarcelluti}\ and\ \citenamefont {Sandin}(2011)}]{CIARCELLUTI201119}%
  \BibitemOpen
  \bibfield  {author} {\bibinfo {author} {\bibfnamefont {P.}~\bibnamefont {Ciarcelluti}}\ and\ \bibinfo {author} {\bibfnamefont {F.}~\bibnamefont {Sandin}},\ }\href {https://doi.org/https://doi.org/10.1016/j.physletb.2010.11.021} {\bibfield  {journal} {\bibinfo  {journal} {Physics Letters B}\ }\textbf {\bibinfo {volume} {695}},\ \bibinfo {pages} {19} (\bibinfo {year} {2011})}\BibitemShut {NoStop}%
\bibitem [{\citenamefont {Freese}\ and\ \citenamefont {Winkler}(2023)}]{PhysRevD.107.083522}%
  \BibitemOpen
  \bibfield  {author} {\bibinfo {author} {\bibfnamefont {K.}~\bibnamefont {Freese}}\ and\ \bibinfo {author} {\bibfnamefont {M.~W.}\ \bibnamefont {Winkler}},\ }\href {https://doi.org/10.1103/PhysRevD.107.083522} {\bibfield  {journal} {\bibinfo  {journal} {Phys. Rev. D}\ }\textbf {\bibinfo {volume} {107}},\ \bibinfo {pages} {083522} (\bibinfo {year} {2023})}\BibitemShut {NoStop}%
\bibitem [{\citenamefont {Kaplan}\ \emph {et~al.}(2009)\citenamefont {Kaplan}, \citenamefont {Luty},\ and\ \citenamefont {Zurek}}]{PhysRevD.79.115016}%
  \BibitemOpen
  \bibfield  {author} {\bibinfo {author} {\bibfnamefont {D.~E.}\ \bibnamefont {Kaplan}}, \bibinfo {author} {\bibfnamefont {M.~A.}\ \bibnamefont {Luty}},\ and\ \bibinfo {author} {\bibfnamefont {K.~M.}\ \bibnamefont {Zurek}},\ }\href {https://doi.org/10.1103/PhysRevD.79.115016} {\bibfield  {journal} {\bibinfo  {journal} {Phys. Rev. D}\ }\textbf {\bibinfo {volume} {79}},\ \bibinfo {pages} {115016} (\bibinfo {year} {2009})}\BibitemShut {NoStop}%
\bibitem [{\citenamefont {Shelton}\ and\ \citenamefont {Zurek}(2010)}]{PhysRevD.82.123512}%
  \BibitemOpen
  \bibfield  {author} {\bibinfo {author} {\bibfnamefont {J.}~\bibnamefont {Shelton}}\ and\ \bibinfo {author} {\bibfnamefont {K.~M.}\ \bibnamefont {Zurek}},\ }\href {https://doi.org/10.1103/PhysRevD.82.123512} {\bibfield  {journal} {\bibinfo  {journal} {Phys. Rev. D}\ }\textbf {\bibinfo {volume} {82}},\ \bibinfo {pages} {123512} (\bibinfo {year} {2010})}\BibitemShut {NoStop}%
\bibitem [{\citenamefont {Giangrandi}\ \emph {et~al.}(2023)\citenamefont {Giangrandi}, \citenamefont {Sagun}, \citenamefont {Ivanytskyi}, \citenamefont {Providência},\ and\ \citenamefont {Dietrich}}]{Giangrandi_2023}%
  \BibitemOpen
  \bibfield  {author} {\bibinfo {author} {\bibfnamefont {E.}~\bibnamefont {Giangrandi}}, \bibinfo {author} {\bibfnamefont {V.}~\bibnamefont {Sagun}}, \bibinfo {author} {\bibfnamefont {O.}~\bibnamefont {Ivanytskyi}}, \bibinfo {author} {\bibfnamefont {C.}~\bibnamefont {Providência}},\ and\ \bibinfo {author} {\bibfnamefont {T.}~\bibnamefont {Dietrich}},\ }\href {https://doi.org/10.3847/1538-4357/ace104} {\bibfield  {journal} {\bibinfo  {journal} {The Astrophysical Journal}\ }\textbf {\bibinfo {volume} {953}},\ \bibinfo {pages} {115} (\bibinfo {year} {2023})}\BibitemShut {NoStop}%
\bibitem [{\citenamefont {Acevedo}\ \emph {et~al.}(2021)\citenamefont {Acevedo}, \citenamefont {Bramante},\ and\ \citenamefont {Goodman}}]{PhysRevD.103.123022}%
  \BibitemOpen
  \bibfield  {author} {\bibinfo {author} {\bibfnamefont {J.~F.}\ \bibnamefont {Acevedo}}, \bibinfo {author} {\bibfnamefont {J.}~\bibnamefont {Bramante}},\ and\ \bibinfo {author} {\bibfnamefont {A.}~\bibnamefont {Goodman}},\ }\href {https://doi.org/10.1103/PhysRevD.103.123022} {\bibfield  {journal} {\bibinfo  {journal} {Phys. Rev. D}\ }\textbf {\bibinfo {volume} {103}},\ \bibinfo {pages} {123022} (\bibinfo {year} {2021})}\BibitemShut {NoStop}%
\bibitem [{\citenamefont {Fuller}\ and\ \citenamefont {Ott}(2015)}]{10.1093/mnrasl/slv049}%
  \BibitemOpen
  \bibfield  {author} {\bibinfo {author} {\bibfnamefont {J.}~\bibnamefont {Fuller}}\ and\ \bibinfo {author} {\bibfnamefont {C.~D.}\ \bibnamefont {Ott}},\ }\href {https://doi.org/10.1093/mnrasl/slv049} {\bibfield  {journal} {\bibinfo  {journal} {Monthly Notices of the Royal Astronomical Society: Letters}\ }\textbf {\bibinfo {volume} {450}},\ \bibinfo {pages} {L71} (\bibinfo {year} {2015})}\BibitemShut {NoStop}%
\bibitem [{\citenamefont {Sedrakian}(2019)}]{PhysRevD.99.043011}%
  \BibitemOpen
  \bibfield  {author} {\bibinfo {author} {\bibfnamefont {A.}~\bibnamefont {Sedrakian}},\ }\href {https://doi.org/10.1103/PhysRevD.99.043011} {\bibfield  {journal} {\bibinfo  {journal} {Phys. Rev. D}\ }\textbf {\bibinfo {volume} {99}},\ \bibinfo {pages} {043011} (\bibinfo {year} {2019})}\BibitemShut {NoStop}%
\bibitem [{\citenamefont {Rafiei~Karkevandi}\ \emph {et~al.}(2022)\citenamefont {Rafiei~Karkevandi}, \citenamefont {Shakeri}, \citenamefont {Sagun},\ and\ \citenamefont {Ivanytskyi}}]{PhysRevD.105.023001}%
  \BibitemOpen
  \bibfield  {author} {\bibinfo {author} {\bibfnamefont {D.}~\bibnamefont {Rafiei~Karkevandi}}, \bibinfo {author} {\bibfnamefont {S.}~\bibnamefont {Shakeri}}, \bibinfo {author} {\bibfnamefont {V.}~\bibnamefont {Sagun}},\ and\ \bibinfo {author} {\bibfnamefont {O.}~\bibnamefont {Ivanytskyi}},\ }\href {https://doi.org/10.1103/PhysRevD.105.023001} {\bibfield  {journal} {\bibinfo  {journal} {Phys. Rev. D}\ }\textbf {\bibinfo {volume} {105}},\ \bibinfo {pages} {023001} (\bibinfo {year} {2022})}\BibitemShut {NoStop}%
\bibitem [{\citenamefont {Nelson}\ \emph {et~al.}(2019)\citenamefont {Nelson}, \citenamefont {Reddy},\ and\ \citenamefont {Zhou}}]{Nelson_2019}%
  \BibitemOpen
  \bibfield  {author} {\bibinfo {author} {\bibfnamefont {A.~E.}\ \bibnamefont {Nelson}}, \bibinfo {author} {\bibfnamefont {S.}~\bibnamefont {Reddy}},\ and\ \bibinfo {author} {\bibfnamefont {D.}~\bibnamefont {Zhou}},\ }\href {https://doi.org/10.1088/1475-7516/2019/07/012} {\bibfield  {journal} {\bibinfo  {journal} {Journal of Cosmology and Astroparticle Physics}\ }\textbf {\bibinfo {volume} {2019}}\bibinfo  {number} { (07)},\ \bibinfo {pages} {012}}\BibitemShut {NoStop}%
\bibitem [{\citenamefont {Ellis}\ \emph {et~al.}(2018{\natexlab{a}})\citenamefont {Ellis}, \citenamefont {H\"utsi}, \citenamefont {Kannike}, \citenamefont {Marzola}, \citenamefont {Raidal},\ and\ \citenamefont {Vaskonen}}]{PhysRevD.97.123007}%
  \BibitemOpen
\bibfield  {number} {  }\bibfield  {author} {\bibinfo {author} {\bibfnamefont {J.}~\bibnamefont {Ellis}}, \bibinfo {author} {\bibfnamefont {G.}~\bibnamefont {H\"utsi}}, \bibinfo {author} {\bibfnamefont {K.}~\bibnamefont {Kannike}}, \bibinfo {author} {\bibfnamefont {L.}~\bibnamefont {Marzola}}, \bibinfo {author} {\bibfnamefont {M.}~\bibnamefont {Raidal}},\ and\ \bibinfo {author} {\bibfnamefont {V.}~\bibnamefont {Vaskonen}},\ }\href {https://doi.org/10.1103/PhysRevD.97.123007} {\bibfield  {journal} {\bibinfo  {journal} {Phys. Rev. D}\ }\textbf {\bibinfo {volume} {97}},\ \bibinfo {pages} {123007} (\bibinfo {year} {2018}{\natexlab{a}})}\BibitemShut {NoStop}%
\bibitem [{\citenamefont {Konstantinou}(2024)}]{Konstantinou_2024}%
  \BibitemOpen
  \bibfield  {author} {\bibinfo {author} {\bibfnamefont {A.}~\bibnamefont {Konstantinou}},\ }\href {https://doi.org/10.3847/1538-4357/ad4701} {\bibfield  {journal} {\bibinfo  {journal} {The Astrophysical Journal}\ }\textbf {\bibinfo {volume} {968}},\ \bibinfo {pages} {83} (\bibinfo {year} {2024})}\BibitemShut {NoStop}%
\bibitem [{\citenamefont {Ellis}\ \emph {et~al.}(2018{\natexlab{b}})\citenamefont {Ellis}, \citenamefont {Hektor}, \citenamefont {H\"{u}tsi}, \citenamefont {Kannike}, \citenamefont {Marzola}, \citenamefont {Raidal},\ and\ \citenamefont {Vaskonen}}]{ELLIS2018607}%
  \BibitemOpen
  \bibfield  {author} {\bibinfo {author} {\bibfnamefont {J.}~\bibnamefont {Ellis}}, \bibinfo {author} {\bibfnamefont {A.}~\bibnamefont {Hektor}}, \bibinfo {author} {\bibfnamefont {G.}~\bibnamefont {H\"{u}tsi}}, \bibinfo {author} {\bibfnamefont {K.}~\bibnamefont {Kannike}}, \bibinfo {author} {\bibfnamefont {L.}~\bibnamefont {Marzola}}, \bibinfo {author} {\bibfnamefont {M.}~\bibnamefont {Raidal}},\ and\ \bibinfo {author} {\bibfnamefont {V.}~\bibnamefont {Vaskonen}},\ }\href {https://doi.org/https://doi.org/10.1016/j.physletb.2018.04.048} {\bibfield  {journal} {\bibinfo  {journal} {Physics Letters B}\ }\textbf {\bibinfo {volume} {781}},\ \bibinfo {pages} {607} (\bibinfo {year} {2018}{\natexlab{b}})}\BibitemShut {NoStop}%
\bibitem [{\citenamefont {Leung}\ \emph {et~al.}(2011)\citenamefont {Leung}, \citenamefont {Chu},\ and\ \citenamefont {Lin}}]{PhysRevD.84.107301}%
  \BibitemOpen
  \bibfield  {author} {\bibinfo {author} {\bibfnamefont {S.-C.}\ \bibnamefont {Leung}}, \bibinfo {author} {\bibfnamefont {M.-C.}\ \bibnamefont {Chu}},\ and\ \bibinfo {author} {\bibfnamefont {L.-M.}\ \bibnamefont {Lin}},\ }\href {https://doi.org/10.1103/PhysRevD.84.107301} {\bibfield  {journal} {\bibinfo  {journal} {Phys. Rev. D}\ }\textbf {\bibinfo {volume} {84}},\ \bibinfo {pages} {107301} (\bibinfo {year} {2011})}\BibitemShut {NoStop}%
\bibitem [{\citenamefont {Xiang}\ \emph {et~al.}(2014)\citenamefont {Xiang}, \citenamefont {Jiang}, \citenamefont {Zhang},\ and\ \citenamefont {Yang}}]{PhysRevC.89.025803}%
  \BibitemOpen
  \bibfield  {author} {\bibinfo {author} {\bibfnamefont {Q.-F.}\ \bibnamefont {Xiang}}, \bibinfo {author} {\bibfnamefont {W.-Z.}\ \bibnamefont {Jiang}}, \bibinfo {author} {\bibfnamefont {D.-R.}\ \bibnamefont {Zhang}},\ and\ \bibinfo {author} {\bibfnamefont {R.-Y.}\ \bibnamefont {Yang}},\ }\href {https://doi.org/10.1103/PhysRevC.89.025803} {\bibfield  {journal} {\bibinfo  {journal} {Phys. Rev. C}\ }\textbf {\bibinfo {volume} {89}},\ \bibinfo {pages} {025803} (\bibinfo {year} {2014})}\BibitemShut {NoStop}%
\bibitem [{\citenamefont {Panotopoulos}\ and\ \citenamefont {Lopes}(2017)}]{PhysRevD.96.083004}%
  \BibitemOpen
  \bibfield  {author} {\bibinfo {author} {\bibfnamefont {G.}~\bibnamefont {Panotopoulos}}\ and\ \bibinfo {author} {\bibfnamefont {I.}~\bibnamefont {Lopes}},\ }\href {https://doi.org/10.1103/PhysRevD.96.083004} {\bibfield  {journal} {\bibinfo  {journal} {Phys. Rev. D}\ }\textbf {\bibinfo {volume} {96}},\ \bibinfo {pages} {083004} (\bibinfo {year} {2017})}\BibitemShut {NoStop}%
\bibitem [{\citenamefont {Das}\ \emph {et~al.}(2019)\citenamefont {Das}, \citenamefont {Malik},\ and\ \citenamefont {Nayak}}]{PhysRevD.99.043016}%
  \BibitemOpen
  \bibfield  {author} {\bibinfo {author} {\bibfnamefont {A.}~\bibnamefont {Das}}, \bibinfo {author} {\bibfnamefont {T.}~\bibnamefont {Malik}},\ and\ \bibinfo {author} {\bibfnamefont {A.~C.}\ \bibnamefont {Nayak}},\ }\href {https://doi.org/10.1103/PhysRevD.99.043016} {\bibfield  {journal} {\bibinfo  {journal} {Phys. Rev. D}\ }\textbf {\bibinfo {volume} {99}},\ \bibinfo {pages} {043016} (\bibinfo {year} {2019})}\BibitemShut {NoStop}%
\bibitem [{\citenamefont {Flores}\ \emph {et~al.}(2024)\citenamefont {Flores}, \citenamefont {Lenzi}, \citenamefont {Dutra}, \citenamefont {Louren\ifmmode~\mbox{\c{c}}\else \c{c}\fi{}o},\ and\ \citenamefont {Arba\~nil}}]{PhysRevD.109.083021}%
  \BibitemOpen
  \bibfield  {author} {\bibinfo {author} {\bibfnamefont {C.~V.}\ \bibnamefont {Flores}}, \bibinfo {author} {\bibfnamefont {C.~H.}\ \bibnamefont {Lenzi}}, \bibinfo {author} {\bibfnamefont {M.}~\bibnamefont {Dutra}}, \bibinfo {author} {\bibfnamefont {O.}~\bibnamefont {Louren\ifmmode~\mbox{\c{c}}\else \c{c}\fi{}o}},\ and\ \bibinfo {author} {\bibfnamefont {J.~D.~V.}\ \bibnamefont {Arba\~nil}},\ }\href {https://doi.org/10.1103/PhysRevD.109.083021} {\bibfield  {journal} {\bibinfo  {journal} {Phys. Rev. D}\ }\textbf {\bibinfo {volume} {109}},\ \bibinfo {pages} {083021} (\bibinfo {year} {2024})}\BibitemShut {NoStop}%
\bibitem [{\citenamefont {Routaray}\ \emph {et~al.}(2023)\citenamefont {Routaray}, \citenamefont {Mohanty}, \citenamefont {Das}, \citenamefont {Ghosh}, \citenamefont {Kalita}, \citenamefont {Parmar},\ and\ \citenamefont {Kumar}}]{Routaray_2023}%
  \BibitemOpen
  \bibfield  {author} {\bibinfo {author} {\bibfnamefont {P.}~\bibnamefont {Routaray}}, \bibinfo {author} {\bibfnamefont {S.~R.}\ \bibnamefont {Mohanty}}, \bibinfo {author} {\bibfnamefont {H.}~\bibnamefont {Das}}, \bibinfo {author} {\bibfnamefont {S.}~\bibnamefont {Ghosh}}, \bibinfo {author} {\bibfnamefont {P.}~\bibnamefont {Kalita}}, \bibinfo {author} {\bibfnamefont {V.}~\bibnamefont {Parmar}},\ and\ \bibinfo {author} {\bibfnamefont {B.}~\bibnamefont {Kumar}},\ }\href {https://doi.org/10.1088/1475-7516/2023/10/073} {\bibfield  {journal} {\bibinfo  {journal} {Journal of Cosmology and Astroparticle Physics}\ }\textbf {\bibinfo {volume} {2023}}\bibinfo  {number} { (10)},\ \bibinfo {pages} {073}}\BibitemShut {NoStop}%
\bibitem [{\citenamefont {Bhat}\ and\ \citenamefont {Paul}(2020)}]{Bhat2020}%
  \BibitemOpen
\bibfield  {number} {  }\bibfield  {author} {\bibinfo {author} {\bibfnamefont {S.~A.}\ \bibnamefont {Bhat}}\ and\ \bibinfo {author} {\bibfnamefont {A.}~\bibnamefont {Paul}},\ }\href {https://doi.org/10.1140/epjc/s10052-020-8072-x} {\bibfield  {journal} {\bibinfo  {journal} {The European Physical Journal C}\ }\textbf {\bibinfo {volume} {80}},\ \bibinfo {pages} {544} (\bibinfo {year} {2020})}\BibitemShut {NoStop}%
\bibitem [{\citenamefont {Kumar}\ \emph {et~al.}(2022)\citenamefont {Kumar}, \citenamefont {Das},\ and\ \citenamefont {Patra}}]{10.1093/mnras/stac1013}%
  \BibitemOpen
  \bibfield  {author} {\bibinfo {author} {\bibfnamefont {A.}~\bibnamefont {Kumar}}, \bibinfo {author} {\bibfnamefont {H.~C.}\ \bibnamefont {Das}},\ and\ \bibinfo {author} {\bibfnamefont {S.~K.}\ \bibnamefont {Patra}},\ }\href {https://doi.org/10.1093/mnras/stac1013} {\bibfield  {journal} {\bibinfo  {journal} {Monthly Notices of the Royal Astronomical Society}\ }\textbf {\bibinfo {volume} {513}},\ \bibinfo {pages} {1820} (\bibinfo {year} {2022})}\BibitemShut {NoStop}%
\bibitem [{\citenamefont {Abac}\ \emph {et~al.}(2023)\citenamefont {Abac}, \citenamefont {Bernido},\ and\ \citenamefont {Esguerra}}]{ABAC2023101185}%
  \BibitemOpen
  \bibfield  {author} {\bibinfo {author} {\bibfnamefont {A.~G.}\ \bibnamefont {Abac}}, \bibinfo {author} {\bibfnamefont {C.~C.}\ \bibnamefont {Bernido}},\ and\ \bibinfo {author} {\bibfnamefont {J.~P.~H.}\ \bibnamefont {Esguerra}},\ }\href {https://doi.org/https://doi.org/10.1016/j.dark.2023.101185} {\bibfield  {journal} {\bibinfo  {journal} {Physics of the Dark Universe}\ }\textbf {\bibinfo {volume} {40}},\ \bibinfo {pages} {101185} (\bibinfo {year} {2023})}\BibitemShut {NoStop}%
\bibitem [{\citenamefont {Das}\ \emph {et~al.}(2021{\natexlab{b}})\citenamefont {Das}, \citenamefont {Kumar},\ and\ \citenamefont {Patra}}]{10.1093/mnras/stab2387}%
  \BibitemOpen
  \bibfield  {author} {\bibinfo {author} {\bibfnamefont {H.~C.}\ \bibnamefont {Das}}, \bibinfo {author} {\bibfnamefont {A.}~\bibnamefont {Kumar}},\ and\ \bibinfo {author} {\bibfnamefont {S.~K.}\ \bibnamefont {Patra}},\ }\href {https://doi.org/10.1093/mnras/stab2387} {\bibfield  {journal} {\bibinfo  {journal} {Monthly Notices of the Royal Astronomical Society}\ }\textbf {\bibinfo {volume} {507}},\ \bibinfo {pages} {4053} (\bibinfo {year} {2021}{\natexlab{b}})}\BibitemShut {NoStop}%
\bibitem [{\citenamefont {Kumar}\ and\ \citenamefont {Sotani}(2024)}]{PhysRevD.110.063001}%
  \BibitemOpen
  \bibfield  {author} {\bibinfo {author} {\bibfnamefont {A.}~\bibnamefont {Kumar}}\ and\ \bibinfo {author} {\bibfnamefont {H.}~\bibnamefont {Sotani}},\ }\href {https://doi.org/10.1103/PhysRevD.110.063001} {\bibfield  {journal} {\bibinfo  {journal} {Phys. Rev. D}\ }\textbf {\bibinfo {volume} {110}},\ \bibinfo {pages} {063001} (\bibinfo {year} {2024})}\BibitemShut {NoStop}%
\bibitem [{\citenamefont {Müller}\ and\ \citenamefont {Serot}(1996)}]{MULLER1996508}%
  \BibitemOpen
  \bibfield  {author} {\bibinfo {author} {\bibfnamefont {H.}~\bibnamefont {Müller}}\ and\ \bibinfo {author} {\bibfnamefont {B.~D.}\ \bibnamefont {Serot}},\ }\href {https://doi.org/https://doi.org/10.1016/0375-9474(96)00187-X} {\bibfield  {journal} {\bibinfo  {journal} {Nuclear Physics A}\ }\textbf {\bibinfo {volume} {606}},\ \bibinfo {pages} {508} (\bibinfo {year} {1996})}\BibitemShut {NoStop}%
\bibitem [{\citenamefont {Boguta}\ and\ \citenamefont {Bodmer}(1977)}]{BOGUTA1977413}%
  \BibitemOpen
  \bibfield  {author} {\bibinfo {author} {\bibfnamefont {J.}~\bibnamefont {Boguta}}\ and\ \bibinfo {author} {\bibfnamefont {A.}~\bibnamefont {Bodmer}},\ }\href {https://doi.org/https://doi.org/10.1016/0375-9474(77)90626-1} {\bibfield  {journal} {\bibinfo  {journal} {Nuclear Physics A}\ }\textbf {\bibinfo {volume} {292}},\ \bibinfo {pages} {413} (\bibinfo {year} {1977})}\BibitemShut {NoStop}%
\bibitem [{\citenamefont {Gambhir}\ and\ \citenamefont {Ring}(1989)}]{Gambhir1989}%
  \BibitemOpen
  \bibfield  {author} {\bibinfo {author} {\bibfnamefont {Y.~K.}\ \bibnamefont {Gambhir}}\ and\ \bibinfo {author} {\bibfnamefont {P.}~\bibnamefont {Ring}},\ }\href {https://doi.org/10.1007/BF02845972} {\bibfield  {journal} {\bibinfo  {journal} {Pramana}\ }\textbf {\bibinfo {volume} {32}},\ \bibinfo {pages} {389} (\bibinfo {year} {1989})}\BibitemShut {NoStop}%
\bibitem [{\citenamefont {Chen}\ and\ \citenamefont {Piekarewicz}(2015)}]{CHEN2015284}%
  \BibitemOpen
  \bibfield  {author} {\bibinfo {author} {\bibfnamefont {W.-C.}\ \bibnamefont {Chen}}\ and\ \bibinfo {author} {\bibfnamefont {J.}~\bibnamefont {Piekarewicz}},\ }\href {https://doi.org/https://doi.org/10.1016/j.physletb.2015.07.020} {\bibfield  {journal} {\bibinfo  {journal} {Physics Letters B}\ }\textbf {\bibinfo {volume} {748}},\ \bibinfo {pages} {284} (\bibinfo {year} {2015})}\BibitemShut {NoStop}%
\bibitem [{\citenamefont {Chen}\ and\ \citenamefont {Piekarewicz}(2014)}]{PhysRevC.90.044305}%
  \BibitemOpen
  \bibfield  {author} {\bibinfo {author} {\bibfnamefont {W.-C.}\ \bibnamefont {Chen}}\ and\ \bibinfo {author} {\bibfnamefont {J.}~\bibnamefont {Piekarewicz}},\ }\href {https://doi.org/10.1103/PhysRevC.90.044305} {\bibfield  {journal} {\bibinfo  {journal} {Phys. Rev. C}\ }\textbf {\bibinfo {volume} {90}},\ \bibinfo {pages} {044305} (\bibinfo {year} {2014})}\BibitemShut {NoStop}%
\bibitem [{\citenamefont {Lalazissis}\ \emph {et~al.}(1997)\citenamefont {Lalazissis}, \citenamefont {K\"onig},\ and\ \citenamefont {Ring}}]{PhysRevC.55.540}%
  \BibitemOpen
  \bibfield  {author} {\bibinfo {author} {\bibfnamefont {G.~A.}\ \bibnamefont {Lalazissis}}, \bibinfo {author} {\bibfnamefont {J.}~\bibnamefont {K\"onig}},\ and\ \bibinfo {author} {\bibfnamefont {P.}~\bibnamefont {Ring}},\ }\href {https://doi.org/10.1103/PhysRevC.55.540} {\bibfield  {journal} {\bibinfo  {journal} {Phys. Rev. C}\ }\textbf {\bibinfo {volume} {55}},\ \bibinfo {pages} {540} (\bibinfo {year} {1997})}\BibitemShut {NoStop}%
\bibitem [{\citenamefont {Fattoyev}\ \emph {et~al.}(2020)\citenamefont {Fattoyev}, \citenamefont {Horowitz}, \citenamefont {Piekarewicz},\ and\ \citenamefont {Reed}}]{PhysRevC.102.065805}%
  \BibitemOpen
  \bibfield  {author} {\bibinfo {author} {\bibfnamefont {F.~J.}\ \bibnamefont {Fattoyev}}, \bibinfo {author} {\bibfnamefont {C.~J.}\ \bibnamefont {Horowitz}}, \bibinfo {author} {\bibfnamefont {J.}~\bibnamefont {Piekarewicz}},\ and\ \bibinfo {author} {\bibfnamefont {B.}~\bibnamefont {Reed}},\ }\href {https://doi.org/10.1103/PhysRevC.102.065805} {\bibfield  {journal} {\bibinfo  {journal} {Phys. Rev. C}\ }\textbf {\bibinfo {volume} {102}},\ \bibinfo {pages} {065805} (\bibinfo {year} {2020})}\BibitemShut {NoStop}%
\bibitem [{\citenamefont {Kumar}\ \emph {et~al.}(2018)\citenamefont {Kumar}, \citenamefont {Patra},\ and\ \citenamefont {Agrawal}}]{PhysRevC.97.045806}%
  \BibitemOpen
  \bibfield  {author} {\bibinfo {author} {\bibfnamefont {B.}~\bibnamefont {Kumar}}, \bibinfo {author} {\bibfnamefont {S.~K.}\ \bibnamefont {Patra}},\ and\ \bibinfo {author} {\bibfnamefont {B.~K.}\ \bibnamefont {Agrawal}},\ }\href {https://doi.org/10.1103/PhysRevC.97.045806} {\bibfield  {journal} {\bibinfo  {journal} {Phys. Rev. C}\ }\textbf {\bibinfo {volume} {97}},\ \bibinfo {pages} {045806} (\bibinfo {year} {2018})}\BibitemShut {NoStop}%
\bibitem [{\citenamefont {Walecka}(1974)}]{WALECKA1974491}%
  \BibitemOpen
  \bibfield  {author} {\bibinfo {author} {\bibfnamefont {J.}~\bibnamefont {Walecka}},\ }\href {https://doi.org/https://doi.org/10.1016/0003-4916(74)90208-5} {\bibfield  {journal} {\bibinfo  {journal} {Annals of Physics}\ }\textbf {\bibinfo {volume} {83}},\ \bibinfo {pages} {491} (\bibinfo {year} {1974})}\BibitemShut {NoStop}%
\bibitem [{\citenamefont {Kubis}\ and\ \citenamefont {Kutschera}(1997)}]{KUBIS1997191}%
  \BibitemOpen
  \bibfield  {author} {\bibinfo {author} {\bibfnamefont {S.}~\bibnamefont {Kubis}}\ and\ \bibinfo {author} {\bibfnamefont {M.}~\bibnamefont {Kutschera}},\ }\href {https://doi.org/https://doi.org/10.1016/S0370-2693(97)00306-7} {\bibfield  {journal} {\bibinfo  {journal} {Physics Letters B}\ }\textbf {\bibinfo {volume} {399}},\ \bibinfo {pages} {191} (\bibinfo {year} {1997})}\BibitemShut {NoStop}%
\bibitem [{\citenamefont {Ma\ifmmode~\acute{n}\else \'{n}\fi{}ka}\ \emph {et~al.}(2000)\citenamefont {Ma\ifmmode~\acute{n}\else \'{n}\fi{}ka}, \citenamefont {Bednarek},\ and\ \citenamefont {Przyby\l{}a}}]{PhysRevC.62.015802}%
  \BibitemOpen
  \bibfield  {author} {\bibinfo {author} {\bibfnamefont {R.}~\bibnamefont {Ma\ifmmode~\acute{n}\else \'{n}\fi{}ka}}, \bibinfo {author} {\bibfnamefont {I.}~\bibnamefont {Bednarek}},\ and\ \bibinfo {author} {\bibfnamefont {G.}~\bibnamefont {Przyby\l{}a}},\ }\href {https://doi.org/10.1103/PhysRevC.62.015802} {\bibfield  {journal} {\bibinfo  {journal} {Phys. Rev. C}\ }\textbf {\bibinfo {volume} {62}},\ \bibinfo {pages} {015802} (\bibinfo {year} {2000})}\BibitemShut {NoStop}%
\bibitem [{\citenamefont {Kumar}\ \emph {et~al.}(2020)\citenamefont {Kumar}, \citenamefont {Das}, \citenamefont {Biswal}, \citenamefont {Kumar},\ and\ \citenamefont {Patra}}]{Kumar2020}%
  \BibitemOpen
  \bibfield  {author} {\bibinfo {author} {\bibfnamefont {A.}~\bibnamefont {Kumar}}, \bibinfo {author} {\bibfnamefont {H.~C.}\ \bibnamefont {Das}}, \bibinfo {author} {\bibfnamefont {S.~K.}\ \bibnamefont {Biswal}}, \bibinfo {author} {\bibfnamefont {B.}~\bibnamefont {Kumar}},\ and\ \bibinfo {author} {\bibfnamefont {S.~K.}\ \bibnamefont {Patra}},\ }\href {https://doi.org/10.1140/epjc/s10052-020-8353-4} {\bibfield  {journal} {\bibinfo  {journal} {The European Physical Journal C}\ }\textbf {\bibinfo {volume} {80}},\ \bibinfo {pages} {775} (\bibinfo {year} {2020})}\BibitemShut {NoStop}%
\bibitem [{\citenamefont {Cline}\ \emph {et~al.}(2013)\citenamefont {Cline}, \citenamefont {Scott}, \citenamefont {Kainulainen},\ and\ \citenamefont {Weniger}}]{PhysRevD.88.055025}%
  \BibitemOpen
  \bibfield  {author} {\bibinfo {author} {\bibfnamefont {J.~M.}\ \bibnamefont {Cline}}, \bibinfo {author} {\bibfnamefont {P.}~\bibnamefont {Scott}}, \bibinfo {author} {\bibfnamefont {K.}~\bibnamefont {Kainulainen}},\ and\ \bibinfo {author} {\bibfnamefont {C.}~\bibnamefont {Weniger}},\ }\href {https://doi.org/10.1103/PhysRevD.88.055025} {\bibfield  {journal} {\bibinfo  {journal} {Phys. Rev. D}\ }\textbf {\bibinfo {volume} {88}},\ \bibinfo {pages} {055025} (\bibinfo {year} {2013})}\BibitemShut {NoStop}%
\bibitem [{\citenamefont {{Douchin, F.}}\ and\ \citenamefont {{Haensel, P.}}(2001)}]{refId0}%
  \BibitemOpen
  \bibfield  {author} {\bibinfo {author} {\bibnamefont {{Douchin, F.}}}\ and\ \bibinfo {author} {\bibnamefont {{Haensel, P.}}},\ }\href {https://doi.org/10.1051/0004-6361:20011402} {\bibfield  {journal} {\bibinfo  {journal} {A\&A}\ }\textbf {\bibinfo {volume} {380}},\ \bibinfo {pages} {151} (\bibinfo {year} {2001})}\BibitemShut {NoStop}%
\bibitem [{\citenamefont {Riley}\ \emph {et~al.}(2019)\citenamefont {Riley}, \citenamefont {Watts}, \citenamefont {Bogdanov}, \citenamefont {Ray}, \citenamefont {Ludlam}, \citenamefont {Guillot}, \citenamefont {Arzoumanian}, \citenamefont {Baker}, \citenamefont {Bilous}, \citenamefont {Chakrabarty}, \citenamefont {Gendreau}, \citenamefont {Harding}, \citenamefont {Ho}, \citenamefont {Lattimer}, \citenamefont {Morsink},\ and\ \citenamefont {Strohmayer}}]{Riley_2019}%
  \BibitemOpen
  \bibfield  {author} {\bibinfo {author} {\bibfnamefont {T.~E.}\ \bibnamefont {Riley}}, \bibinfo {author} {\bibfnamefont {A.~L.}\ \bibnamefont {Watts}}, \bibinfo {author} {\bibfnamefont {S.}~\bibnamefont {Bogdanov}}, \bibinfo {author} {\bibfnamefont {P.~S.}\ \bibnamefont {Ray}}, \bibinfo {author} {\bibfnamefont {R.~M.}\ \bibnamefont {Ludlam}}, \bibinfo {author} {\bibfnamefont {S.}~\bibnamefont {Guillot}}, \bibinfo {author} {\bibfnamefont {Z.}~\bibnamefont {Arzoumanian}}, \bibinfo {author} {\bibfnamefont {C.~L.}\ \bibnamefont {Baker}}, \bibinfo {author} {\bibfnamefont {A.~V.}\ \bibnamefont {Bilous}}, \bibinfo {author} {\bibfnamefont {D.}~\bibnamefont {Chakrabarty}}, \bibinfo {author} {\bibfnamefont {K.~C.}\ \bibnamefont {Gendreau}}, \bibinfo {author} {\bibfnamefont {A.~K.}\ \bibnamefont {Harding}}, \bibinfo {author} {\bibfnamefont {W.~C.~G.}\ \bibnamefont {Ho}}, \bibinfo {author} {\bibfnamefont {J.~M.}\ \bibnamefont {Lattimer}}, \bibinfo {author} {\bibfnamefont {S.~M.}\ \bibnamefont {Morsink}},\ and\ \bibinfo
  {author} {\bibfnamefont {T.~E.}\ \bibnamefont {Strohmayer}},\ }\href {https://doi.org/10.3847/2041-8213/ab481c} {\bibfield  {journal} {\bibinfo  {journal} {The Astrophysical Journal Letters}\ }\textbf {\bibinfo {volume} {887}},\ \bibinfo {pages} {L21} (\bibinfo {year} {2019})}\BibitemShut {NoStop}%
\bibitem [{\citenamefont {Miller}\ \emph {et~al.}(2019)\citenamefont {Miller}, \citenamefont {Lamb}, \citenamefont {Dittmann}, \citenamefont {Bogdanov}, \citenamefont {Arzoumanian}, \citenamefont {Gendreau}, \citenamefont {Guillot}, \citenamefont {Harding}, \citenamefont {Ho}, \citenamefont {Lattimer}, \citenamefont {Ludlam}, \citenamefont {Mahmoodifar}, \citenamefont {Morsink}, \citenamefont {Ray}, \citenamefont {Strohmayer}, \citenamefont {Wood}, \citenamefont {Enoto}, \citenamefont {Foster}, \citenamefont {Okajima}, \citenamefont {Prigozhin},\ and\ \citenamefont {Soong}}]{Miller_2019}%
  \BibitemOpen
  \bibfield  {author} {\bibinfo {author} {\bibfnamefont {M.~C.}\ \bibnamefont {Miller}}, \bibinfo {author} {\bibfnamefont {F.~K.}\ \bibnamefont {Lamb}}, \bibinfo {author} {\bibfnamefont {A.~J.}\ \bibnamefont {Dittmann}}, \bibinfo {author} {\bibfnamefont {S.}~\bibnamefont {Bogdanov}}, \bibinfo {author} {\bibfnamefont {Z.}~\bibnamefont {Arzoumanian}}, \bibinfo {author} {\bibfnamefont {K.~C.}\ \bibnamefont {Gendreau}}, \bibinfo {author} {\bibfnamefont {S.}~\bibnamefont {Guillot}}, \bibinfo {author} {\bibfnamefont {A.~K.}\ \bibnamefont {Harding}}, \bibinfo {author} {\bibfnamefont {W.~C.~G.}\ \bibnamefont {Ho}}, \bibinfo {author} {\bibfnamefont {J.~M.}\ \bibnamefont {Lattimer}}, \bibinfo {author} {\bibfnamefont {R.~M.}\ \bibnamefont {Ludlam}}, \bibinfo {author} {\bibfnamefont {S.}~\bibnamefont {Mahmoodifar}}, \bibinfo {author} {\bibfnamefont {S.~M.}\ \bibnamefont {Morsink}}, \bibinfo {author} {\bibfnamefont {P.~S.}\ \bibnamefont {Ray}}, \bibinfo {author} {\bibfnamefont {T.~E.}\ \bibnamefont {Strohmayer}}, \bibinfo
  {author} {\bibfnamefont {K.~S.}\ \bibnamefont {Wood}}, \bibinfo {author} {\bibfnamefont {T.}~\bibnamefont {Enoto}}, \bibinfo {author} {\bibfnamefont {R.}~\bibnamefont {Foster}}, \bibinfo {author} {\bibfnamefont {T.}~\bibnamefont {Okajima}}, \bibinfo {author} {\bibfnamefont {G.}~\bibnamefont {Prigozhin}},\ and\ \bibinfo {author} {\bibfnamefont {Y.}~\bibnamefont {Soong}},\ }\href {https://doi.org/10.3847/2041-8213/ab50c5} {\bibfield  {journal} {\bibinfo  {journal} {The Astrophysical Journal Letters}\ }\textbf {\bibinfo {volume} {887}},\ \bibinfo {pages} {L24} (\bibinfo {year} {2019})}\BibitemShut {NoStop}%
\bibitem [{\citenamefont {Riley}\ \emph {et~al.}(2021)\citenamefont {Riley}, \citenamefont {Watts}, \citenamefont {Ray}, \citenamefont {Bogdanov}, \citenamefont {Guillot}, \citenamefont {Morsink}, \citenamefont {Bilous}, \citenamefont {Arzoumanian}, \citenamefont {Choudhury}, \citenamefont {Deneva}, \citenamefont {Gendreau}, \citenamefont {Harding}, \citenamefont {Ho}, \citenamefont {Lattimer}, \citenamefont {Loewenstein}, \citenamefont {Ludlam}, \citenamefont {Markwardt}, \citenamefont {Okajima}, \citenamefont {Prescod-Weinstein}, \citenamefont {Remillard}, \citenamefont {Wolff}, \citenamefont {Fonseca}, \citenamefont {Cromartie}, \citenamefont {Kerr}, \citenamefont {Pennucci}, \citenamefont {Parthasarathy}, \citenamefont {Ransom}, \citenamefont {Stairs}, \citenamefont {Guillemot},\ and\ \citenamefont {Cognard}}]{Riley_2021}%
  \BibitemOpen
  \bibfield  {author} {\bibinfo {author} {\bibfnamefont {T.~E.}\ \bibnamefont {Riley}}, \bibinfo {author} {\bibfnamefont {A.~L.}\ \bibnamefont {Watts}}, \bibinfo {author} {\bibfnamefont {P.~S.}\ \bibnamefont {Ray}}, \bibinfo {author} {\bibfnamefont {S.}~\bibnamefont {Bogdanov}}, \bibinfo {author} {\bibfnamefont {S.}~\bibnamefont {Guillot}}, \bibinfo {author} {\bibfnamefont {S.~M.}\ \bibnamefont {Morsink}}, \bibinfo {author} {\bibfnamefont {A.~V.}\ \bibnamefont {Bilous}}, \bibinfo {author} {\bibfnamefont {Z.}~\bibnamefont {Arzoumanian}}, \bibinfo {author} {\bibfnamefont {D.}~\bibnamefont {Choudhury}}, \bibinfo {author} {\bibfnamefont {J.~S.}\ \bibnamefont {Deneva}}, \bibinfo {author} {\bibfnamefont {K.~C.}\ \bibnamefont {Gendreau}}, \bibinfo {author} {\bibfnamefont {A.~K.}\ \bibnamefont {Harding}}, \bibinfo {author} {\bibfnamefont {W.~C.~G.}\ \bibnamefont {Ho}}, \bibinfo {author} {\bibfnamefont {J.~M.}\ \bibnamefont {Lattimer}}, \bibinfo {author} {\bibfnamefont {M.}~\bibnamefont {Loewenstein}}, \bibinfo
  {author} {\bibfnamefont {R.~M.}\ \bibnamefont {Ludlam}}, \bibinfo {author} {\bibfnamefont {C.~B.}\ \bibnamefont {Markwardt}}, \bibinfo {author} {\bibfnamefont {T.}~\bibnamefont {Okajima}}, \bibinfo {author} {\bibfnamefont {C.}~\bibnamefont {Prescod-Weinstein}}, \bibinfo {author} {\bibfnamefont {R.~A.}\ \bibnamefont {Remillard}}, \bibinfo {author} {\bibfnamefont {M.~T.}\ \bibnamefont {Wolff}}, \bibinfo {author} {\bibfnamefont {E.}~\bibnamefont {Fonseca}}, \bibinfo {author} {\bibfnamefont {H.~T.}\ \bibnamefont {Cromartie}}, \bibinfo {author} {\bibfnamefont {M.}~\bibnamefont {Kerr}}, \bibinfo {author} {\bibfnamefont {T.~T.}\ \bibnamefont {Pennucci}}, \bibinfo {author} {\bibfnamefont {A.}~\bibnamefont {Parthasarathy}}, \bibinfo {author} {\bibfnamefont {S.}~\bibnamefont {Ransom}}, \bibinfo {author} {\bibfnamefont {I.}~\bibnamefont {Stairs}}, \bibinfo {author} {\bibfnamefont {L.}~\bibnamefont {Guillemot}},\ and\ \bibinfo {author} {\bibfnamefont {I.}~\bibnamefont {Cognard}},\ }\href
  {https://doi.org/10.3847/2041-8213/ac0a81} {\bibfield  {journal} {\bibinfo  {journal} {The Astrophysical Journal Letters}\ }\textbf {\bibinfo {volume} {918}},\ \bibinfo {pages} {L27} (\bibinfo {year} {2021})}\BibitemShut {NoStop}%
\bibitem [{\citenamefont {Miller}\ \emph {et~al.}(2021)\citenamefont {Miller}, \citenamefont {Lamb}, \citenamefont {Dittmann}, \citenamefont {Bogdanov}, \citenamefont {Arzoumanian}, \citenamefont {Gendreau}, \citenamefont {Guillot}, \citenamefont {Ho}, \citenamefont {Lattimer}, \citenamefont {Loewenstein}, \citenamefont {Morsink}, \citenamefont {Ray}, \citenamefont {Wolff}, \citenamefont {Baker}, \citenamefont {Cazeau}, \citenamefont {Manthripragada}, \citenamefont {Markwardt}, \citenamefont {Okajima}, \citenamefont {Pollard}, \citenamefont {Cognard}, \citenamefont {Cromartie}, \citenamefont {Fonseca}, \citenamefont {Guillemot}, \citenamefont {Kerr}, \citenamefont {Parthasarathy}, \citenamefont {Pennucci}, \citenamefont {Ransom},\ and\ \citenamefont {Stairs}}]{Miller_2021}%
  \BibitemOpen
  \bibfield  {author} {\bibinfo {author} {\bibfnamefont {M.~C.}\ \bibnamefont {Miller}}, \bibinfo {author} {\bibfnamefont {F.~K.}\ \bibnamefont {Lamb}}, \bibinfo {author} {\bibfnamefont {A.~J.}\ \bibnamefont {Dittmann}}, \bibinfo {author} {\bibfnamefont {S.}~\bibnamefont {Bogdanov}}, \bibinfo {author} {\bibfnamefont {Z.}~\bibnamefont {Arzoumanian}}, \bibinfo {author} {\bibfnamefont {K.~C.}\ \bibnamefont {Gendreau}}, \bibinfo {author} {\bibfnamefont {S.}~\bibnamefont {Guillot}}, \bibinfo {author} {\bibfnamefont {W.~C.~G.}\ \bibnamefont {Ho}}, \bibinfo {author} {\bibfnamefont {J.~M.}\ \bibnamefont {Lattimer}}, \bibinfo {author} {\bibfnamefont {M.}~\bibnamefont {Loewenstein}}, \bibinfo {author} {\bibfnamefont {S.~M.}\ \bibnamefont {Morsink}}, \bibinfo {author} {\bibfnamefont {P.~S.}\ \bibnamefont {Ray}}, \bibinfo {author} {\bibfnamefont {M.~T.}\ \bibnamefont {Wolff}}, \bibinfo {author} {\bibfnamefont {C.~L.}\ \bibnamefont {Baker}}, \bibinfo {author} {\bibfnamefont {T.}~\bibnamefont {Cazeau}}, \bibinfo {author}
  {\bibfnamefont {S.}~\bibnamefont {Manthripragada}}, \bibinfo {author} {\bibfnamefont {C.~B.}\ \bibnamefont {Markwardt}}, \bibinfo {author} {\bibfnamefont {T.}~\bibnamefont {Okajima}}, \bibinfo {author} {\bibfnamefont {S.}~\bibnamefont {Pollard}}, \bibinfo {author} {\bibfnamefont {I.}~\bibnamefont {Cognard}}, \bibinfo {author} {\bibfnamefont {H.~T.}\ \bibnamefont {Cromartie}}, \bibinfo {author} {\bibfnamefont {E.}~\bibnamefont {Fonseca}}, \bibinfo {author} {\bibfnamefont {L.}~\bibnamefont {Guillemot}}, \bibinfo {author} {\bibfnamefont {M.}~\bibnamefont {Kerr}}, \bibinfo {author} {\bibfnamefont {A.}~\bibnamefont {Parthasarathy}}, \bibinfo {author} {\bibfnamefont {T.~T.}\ \bibnamefont {Pennucci}}, \bibinfo {author} {\bibfnamefont {S.}~\bibnamefont {Ransom}},\ and\ \bibinfo {author} {\bibfnamefont {I.}~\bibnamefont {Stairs}},\ }\href {https://doi.org/10.3847/2041-8213/ac089b} {\bibfield  {journal} {\bibinfo  {journal} {The Astrophysical Journal Letters}\ }\textbf {\bibinfo {volume} {918}},\ \bibinfo {pages}
  {L28} (\bibinfo {year} {2021})}\BibitemShut {NoStop}%
\bibitem [{\citenamefont {Blaschke}\ \emph {et~al.}(2020)\citenamefont {Blaschke}, \citenamefont {Ayriyan}, \citenamefont {Alvarez-Castillo},\ and\ \citenamefont {Grigorian}}]{universe6060081}%
  \BibitemOpen
  \bibfield  {author} {\bibinfo {author} {\bibfnamefont {D.}~\bibnamefont {Blaschke}}, \bibinfo {author} {\bibfnamefont {A.}~\bibnamefont {Ayriyan}}, \bibinfo {author} {\bibfnamefont {D.~E.}\ \bibnamefont {Alvarez-Castillo}},\ and\ \bibinfo {author} {\bibfnamefont {H.}~\bibnamefont {Grigorian}},\ }\bibfield  {journal} {\bibinfo  {journal} {Universe}\ }\textbf {\bibinfo {volume} {6}},\ \href {https://doi.org/10.3390/universe6060081} {10.3390/universe6060081} (\bibinfo {year} {2020})\BibitemShut {NoStop}%
\bibitem [{\citenamefont {Fonseca}\ \emph {et~al.}(2021)\citenamefont {Fonseca}, \citenamefont {Cromartie}, \citenamefont {Pennucci}, \citenamefont {Ray}, \citenamefont {Kirichenko}, \citenamefont {Ransom}, \citenamefont {Demorest}, \citenamefont {Stairs}, \citenamefont {Arzoumanian}, \citenamefont {Guillemot}, \citenamefont {Parthasarathy}, \citenamefont {Kerr} \emph {et~al.}}]{Fonseca_2021}%
  \BibitemOpen
  \bibfield  {author} {\bibinfo {author} {\bibfnamefont {E.}~\bibnamefont {Fonseca}}, \bibinfo {author} {\bibfnamefont {H.~T.}\ \bibnamefont {Cromartie}}, \bibinfo {author} {\bibfnamefont {T.~T.}\ \bibnamefont {Pennucci}}, \bibinfo {author} {\bibfnamefont {P.~S.}\ \bibnamefont {Ray}}, \bibinfo {author} {\bibfnamefont {A.~Y.}\ \bibnamefont {Kirichenko}}, \bibinfo {author} {\bibfnamefont {S.~M.}\ \bibnamefont {Ransom}}, \bibinfo {author} {\bibfnamefont {P.~B.}\ \bibnamefont {Demorest}}, \bibinfo {author} {\bibfnamefont {I.~H.}\ \bibnamefont {Stairs}}, \bibinfo {author} {\bibfnamefont {Z.}~\bibnamefont {Arzoumanian}}, \bibinfo {author} {\bibfnamefont {L.}~\bibnamefont {Guillemot}}, \bibinfo {author} {\bibfnamefont {A.}~\bibnamefont {Parthasarathy}}, \bibinfo {author} {\bibfnamefont {M.}~\bibnamefont {Kerr}}, \emph {et~al.},\ }\href {https://doi.org/10.3847/2041-8213/ac03b8} {\bibfield  {journal} {\bibinfo  {journal} {The Astrophysical Journal Letters}\ }\textbf {\bibinfo {volume} {915}},\ \bibinfo {pages} {L12}
  (\bibinfo {year} {2021})}\BibitemShut {NoStop}%
\bibitem [{\citenamefont {Abbott}\ \emph {et~al.}(2018)\citenamefont {Abbott}, \citenamefont {Abbott}, \citenamefont {Abbott}, \citenamefont {Acernese}, \citenamefont {Ackley}, \citenamefont {Adams},\ and\ \citenamefont {et. al. (The LIGO Scientific Collaboration \text{and}~the Virgo~Collaboration)}}]{PhysRevLett.121.161101}%
  \BibitemOpen
  \bibfield  {author} {\bibinfo {author} {\bibfnamefont {B.~P.}\ \bibnamefont {Abbott}}, \bibinfo {author} {\bibfnamefont {R.}~\bibnamefont {Abbott}}, \bibinfo {author} {\bibfnamefont {T.~D.}\ \bibnamefont {Abbott}}, \bibinfo {author} {\bibfnamefont {F.}~\bibnamefont {Acernese}}, \bibinfo {author} {\bibfnamefont {K.}~\bibnamefont {Ackley}}, \bibinfo {author} {\bibfnamefont {C.}~\bibnamefont {Adams}},\ and\ \bibinfo {author} {\bibnamefont {et. al. (The LIGO Scientific Collaboration \text{and}~the Virgo~Collaboration)}},\ }\href {https://doi.org/10.1103/PhysRevLett.121.161101} {\bibfield  {journal} {\bibinfo  {journal} {Phys. Rev. Lett.}\ }\textbf {\bibinfo {volume} {121}},\ \bibinfo {pages} {161101} (\bibinfo {year} {2018})}\BibitemShut {NoStop}%
\end{thebibliography}%
\end{document}